%% file: mnras_template.tex
\documentclass[fleqn,usenatbib]{mnras}
\usepackage{newtxtext,newtxmath}

\usepackage[T1]{fontenc}
\usepackage{ae,aecompl}
\usepackage[dvipsnames]{xcolor}

\usepackage{graphicx}	
\usepackage{amsmath}	
\usepackage{amssymb}	

\newcommand\comment[1]{}
\newcommand\inorig[1]{}

\title[Morphological components in EAGLE]{The star formation rate and stellar content contributions of morphological components in the EAGLE simulations}

\author[J. W. Trayford]{James W. Trayford$^{1,2}$\thanks{E-mail: trayford@strw.leidenuniv.nl}, Carlos S. Frenk$^{1}$, Tom Theuns$^{1}$, Joop Schaye$^{2}$, \newauthor 
Camila Correa$^{2}$
\\
$^{1}$Institute for Computational Cosmology, Ogden Centre - West, Durham University, South Road, Durham DH1 3LE, England \\
$^{2}$Leiden Observatory, Leiden University, P.O. Box 9513, 2300 RA Leiden, The Netherlands
}

\date{Accepted XXX. Received YYY; in original form ZZZ}

\pubyear{2018}

\begin{document}
\label{firstpage}
\pagerange{\pageref{firstpage}--\pageref{lastpage}}
\maketitle

\begin{abstract}
The Hubble sequence provides a useful classification of galaxy
morphology at low redshift.  However, morphologies are not static, but
rather evolve as the growth of structure proceeds through mergers,
accretion and secular processes. We investigate how kinematically
defined disc and spheroidal structures form and evolve in the EAGLE
hydrodynamic simulation of galaxy formation. At high redshift most
galaxies of all masses are asymmetric. By redshift $z\simeq 1.5$ the
Hubble sequence is established and after this time most of the stellar
mass is in spheroids whose contribution to the stellar mass budget continues to rise to the present day. The stellar mass
fraction in discs peaks at $z\simeq 0.5$ but overall remains
subdominant at all times although discs contribute most of the stellar
mass in systems of mass $M^*\sim 10^{10.5}{\rm M_\odot}$ at $z \leq
1.5$. Star formation occurs predominantly in disc structures
throughout most of cosmic time but morphological transformations
rearrange stars, thus establishing the low-redshift morphological mix. 
Morphological transformations are common and we quantify the
rates at which they occur. The rate of growth of spheroids decreases
at $z < 2$ while the rate of decay of discs remains roughly constant
at $z < 1$.  Finally, we find that the prograde component of galaxies
becomes increasingly dynamically cold with time.

\end{abstract}



\input{Intro}
\input{Simulation}
\input{Morphologies}
\input{MacroMorph}
\input{TrackMorph}
\input{MicroMorph}



\section*{Acknowledgements}

JT acknowledges Adrien Thob, Bart Clauwens, Marijn Franx and Claudia Lagos for useful discussions that helped to guide this work. This study made use of the DiRAC Data Centric system at Durham University, operated by the Institute for Computational Cosmology on behalf of the STFC DiRAC HPC Facility ({\tt www.dirac.ac.uk}). This equipment was funded by BIS National E-infrastructure capital grant ST/K00042X/1, STFC  capital  grant  ST/H008519/1,  and  STFC  DiRAC Operations  grant  ST/K003267/1  and  Durham  University. DiRAC is part of the National E-Infrastructure. 

\bibliographystyle{mnras}
\bibliography{references}


\input{Appendices}

\label{lastpage}
\end{document}

%% file: Intro.tex
\section{Introduction}

Although observed galaxies exhibit broad diversity in appearance, taxonomy of galaxies into morphological classes is useful to study common features of their formation and evolution. The Hubble sequence (HS) endures as the primary classification scheme for galaxies from the earliest days of extragalactic observation. In its most basic form the HS arranges galaxies by their light profiles, from smooth and spheroidal \textit{`early'}-types to increasingly disc-dominated and structured \textit{`late'}-types. Galaxy classification relies on visual inspection, so is inherently subjective and can be arduous to perform for large datasets. Parametric light profile fitting methods have been developed to quantify morphology, and are simpler to automate. These methods often use fixed profile shapes to represent discs and spheroids or assume generalised \citet{Sersic68} profiles, and have been applied to significant galaxy samples \citep[e.g.][]{Driver06, Benson07, Kelvin12, Haussler13} providing useful proxies for Hubble type \citep{Kelvin12, Vulcani14}. In addition, citizen science projects such as Galaxy Zoo \citep{Lintott08} use volunteers to obtain multiple independent classifications for large sets ($\sim 10^5$) of galaxy images, enabling a statistical approach to visual classification. 

Alongside the common structural features used to identify early and late archetypes, morphology is also observed to correlate with a number of other properties. Generally a higher fraction of classical early types are identified at high stellar mass or high luminosities \citep[e.g.][]{Benson07, Kelvin14, Moffett16}. For a given stellar mass, the colour-morphology plane also displays bimodality, with blue late-types and red early-types \citep[e.g.][]{Larson80, Strateva01, Baldry04}, indicative of underlying trends between HS \textit{lateness} and specific star formation rate \citep[e.g.][]{Kennicutt83, Kauffmann03, Bluck14, Whitaker15}. Galaxy morphologies have also long been associated with the environments in which they reside, with more early types present in denser environments and vice versa \citep{Oemler74, Dressler80}. Statistics afforded by larger samples show that the morphology-colour and morphology-density relations are distinct \citep[e.g.][]{Bamford09, Skibba09}. 

Physical insight into these systematic trends, and the emergence of Hubble-type galaxies generally, requires an understanding of the underlying kinematic structures of galaxies. Contemporary integral field unit (IFU) studies allow galaxies to be classified by their kinematics \citep[e.g.][]{Emsellem07,Emsellem11}. The kinematic and Hubble classes are far from coincident, but correlate once the anomalous \textit{slow-rotator} class is accounted for \citep{Krajnovic13, Cortese16}. 

The implicit physical picture of HS galaxies as some composite of discs supported by ordered rotation and spheroids supported by stellar dynamical pressure is clearly a simplification, but one that is useful to relate to theoretical considerations \citep[e.g.][]{Abadi03, Parry09}. It provides a framework for understanding morphological evolution as the build-up and interaction of spheroid and disc structures in galaxies. The mass ratio of these two kinematic structures, from completely spheroidal to completely disky, can then be thought of as a measure of the \textit{lateness} of a galaxy along the HS. This decomposition is analogous to measuring the relative photometric contributions of bulges and discs, derived by decomposing galaxy light profiles \citep[e.g.][]{Benson07}.

The association between discs and star formation observed at $z\sim 0$ hints at the importance of these structures for galaxy formation. Generally, discs are considered to result from angular momentum conservation of gas collapsing into dark matter halos \citep[e.g.][]{Fall80, Mo98, Cole00}. Other mechanisms have been hypothesised to enhance disc formation, such as collimated accretion onto galaxies through cosmic filaments \citep{Dekel09, Brooks09} and gas rich mergers \citep{Robertson06, Naab06, Lagos18}. The relationship between the disc and spheroid in late types is also complex, with observed trends between the rotational properties of discs and the prominence of a central bulge \citep[e.g.][]{Obreschkow14}. 

The formation of spheroids has been attributed variously to monolithic collapse of gas and subsequent star formation at high redshift \citep{Eggen67}, tidal interactions and mergers \citep{Toomre77}, disc instabilities \citep[e.g.][]{Parry09, DeLucia11}, and misaligned accretion events \citep[e.g.][]{Sales12, Bett12}.  The importance of each process determines whether spheroids grow primarily through internal star formation, or by subsuming stars formed in other structures, particularly discs. 

Despite the rich theory surrounding the co-evolution of discs and spheroids, considering these structures alone is a false dichotomy. Some of the most local starburst galaxies reveal highly disturbed or peculiar morphologies \citep[e.g. Arp220,][]{Aalto09} or are at least an uncomfortable fit to the HS \citep[e.g. M82,][]{Divakara09}. Disturbed morphologies are considered transient, and typically associated with significant merger or accretion events. Disturbance in the light profile can be quantified by non-parametric measures of morphology, such as shape asymmetry, and are demonstrated to correlate well with starburst activity \citep{Pawlik16}. Many such non-parametric diagnostics for morphology have been developed \citep[e.g.][]{Kent85, Abraham94, Conselice00, Lotz04}, and hint at salient structural properties of galaxies beyond the one-dimensional picture provided by disc to spheroid ratio or Hubble class alone. 

HS nonconformity may be relatively rare at low redshifts, but the importance of disturbed structures increases with lookback time. Indeed, exotic morphological types without local analogues are found in abundance at high redshift \citep{Cowie95, vandenBergh96, Elmegreen04}. Higher merger rates and gas fractions of galaxies in the early universe likely contribute to this greater prevalence of peculiar morphologies \citep{Abraham99, Abraham01}. However the HS is still in evidence at high redshift, with discs and central spheroids observed out to $z \lesssim 3$  \citep[e.g.][]{Tachella15, Swinbank15}. The relative demographics of HS and peculiar galaxies in the early universe are hard to pin down because they are typically observed in the rest-frame UV, and are thus biased towards younger stellar populations. Image degradation and selection biases also typically increase with redshift. 

Some questions that arise from the observational and theoretical work described above are: \begin{itemize}
\item When do HS galaxies come to dominate the galaxy population over peculiar morphologies?
\item What proportion of the total star formation budget over cosmic time is contributed by discs?
\item Do stellar spheroids subsume stars from other structures, or are they mainly built through \textit{in-situ} star formation?
\end{itemize}

Simulations of galaxy formation are valuable tools in addressing these questions, as they can model the complex and competing mechanisms that affect the evolution of galaxy morphologies. The caveat is that one must consider both how well a simulation represents reality, and how the physical structure of galaxies in the simulation maps onto observable properties. As a result, it is necessary to understand the \textit{physical} morphologies (or structure) of simulated galaxies, and how those correspond to the \textit{classical} morphologies that are observed.

Early modelling work in the $\Lambda$CDM paradigm attributed the structure and sizes of galaxies to the spin of their host dark matter halos, assuming they share their angular momentum with the baryons and that this is conserved during galaxy formation \citep{Fall80, Fall83, Mo98, Cole00}. Hydrodynamical simulations initially struggled with \textit{overcooling}: runaway star formation at high redshift would yield a surfeit of massive, compact and bulge-dominated systems \citep[e.g.][]{Katz91, Navarro97, Zavala08, Crain09}. This highlighted the crucial role of efficient feedback in regulating star formation \citep[e.g.][]{Benson03}, particularly by inhibiting excessive bulge formation via the removal of low angular momentum gas from galaxies \citep[e.g.][]{SommerLarsen99, Binney01, Scannapieco08, Governato09, Brook11, Agertz11}. The morphologies of galaxies have been found to be highly sensitive to the strength and implementation of this feedback \citep[e.g.][]{Okamoto05,Scannapieco09,Sales10}.

Simulations of zoomed regions or single halos allow the effects of baryonic physics on morphology to be explored by performing numerical experiments. Case studies exploring the kinematics of late types \citep[e.g.][]{Abadi03}, early types \citep[e.g][]{Meza03} and merger remnants \citep[e.g.][]{Naab06, Robertson06} elucidate the interplay of star formation and feedback. Cosmological zoom simulations are now able to produce galactic discs in a $\Lambda$CDM context  with reasonable disc-to-spheroid ratios \citep[e.g.][]{Governato07, Scannapieco09, Guedes11, Aumer13}. These simulations exhibit complex phase space structure \citep[e.g.][]{GarrisonKimmel17, Grand17, ElBadry18}, and emulate realistic visual morphologies \citep[e.g.][]{Guedes11,Stinson13,Hopkins17}. These may go beyond the simple dichotomy of disc and spheroid, forming distinct psuedobulges, stellar halos and bars  \citep[e.g.][]{Guedes13, Pillepich14, Okamoto15} and detailed vertical structure of discs \citep[e.g.][]{Brook12, Ma17, Navarro17}. 

The high resolution achieved by zoom simulations is currently unfeasible in cosmological volumes. However, large volume simulations have developed feedback and star formation implementations that can effectively overcome \textit{overcooling} and reproduce the observed galaxy stellar mass functions \citep[][]{Vogelsberger14, Schaye15, Dave16, Pillepich18}. These simulations can also produce morphologies across the HS, with a number of studies corroborating the theoretical link between angular momentum of halos and galaxy morphology \citep[e.g.][]{Dubois14, Genel15, Zavala16, RodriguezGomez17, Lagos17b}. However, morphologies are also sensitive to assembly history and feedback, with some authors finding little residual connection with halo spin  \citep{Scannapieco09,Sales12}.

In this work we use the EAGLE suite of cosmological, hydrodynamical simulations \citep{Schaye15, Crain15, McAlpine16} to address the three questions posed above. EAGLE reproduces observed gas properties at low redshift \citep{Lagos15, Bahe16}, galaxy colours \citep{Trayford15, Trayford17} as well as the evolution of the galaxy stellar mass function and galaxy sizes \citep{Furlong15, Furlong17}. Given the resolution of EAGLE, structures such as stellar bars may only be marginally resolved \citep{Algorry17} and discs may be artificially thicker than observed \citep{Llambay18}. In this study we merely distinguish between the rotation-supported (\textit{disc}) and pressure-supported (\textit{spheroid}) components of HS galaxies, and separate the HS from non HS galaxies based on the symmetry of the stellar mass distribution (\textit{asymmetric} systems).  

This work expands upon previous analysis of morphology in EAGLE galaxies, particularly \citet{Zavala16}, \citet{Correa17} and \citet{Clauwens17}. These studies quantify morphology using stellar kinematics, exploring the build up of discs and spheroids \citep{Zavala16,Clauwens17} and the link between physical morphology and observables such as galaxy colour \citep{Correa17}. Here, analysis is purely in the physical domain, with no direct comparison to observations. The aim of this study is to understand the \textit{physical} structure of EAGLE galaxies and characterise the generic processes that determine their evolution. This foundational understanding can provide insight for future studies validating the \textit{visual} morphologies of EAGLE  \citep[via mock observations, e.g.][]{Camps16, Trayford17}\footnote{Complementary \textit{visual} morphology studies are needed to understand how well the simulation reproduces data, and the observational effects \citep{Scannapieco10, Snyder15}.}. Understanding the link between \textit{visual} and \textit{physical} morphology in simulations may provide insight into their connection in the data.

We first describe the EAGLE simulations and our galaxy definition in Section \ref{sec:sims}. We go on to describe various metrics for morphology, and characterise how these apply to our EAGLE galaxy selection in Section \ref{sec:morphs}. Here, we also describe our primary method for separating HS from non-HS (\textit{peculiar}) galaxies, and further decompose HS galaxies into disc and spheroid components based on kinematics.

We then separate stellar mass into \textit{disc}, \textit{spheroid} and \textit{asymmetric} structures, and analyse the mass contribution of each as a function of stellar mass and redshift in Section \ref{sec:macro}. For Section \ref{sec:tree} we  use the simulation merger trees to investigate how the ratio of disc to spheroid evolves in individual galaxies, and the role of mergers in transforming disc-dominated systems into spheroidal systems. In Section \ref{sec:rates} we directly compute the star formation rate contribution of \textit{disc}, \textit{spheroid} and \textit{asymmetric} structures, and use this in combination with the stellar decomposition to show how star formation and transformational process lead to the growth and decay of these structures. We assess our adopted kinematic definitions of disc and spheroid, and the evolving kinematics of individual particles in Section \ref{sec:micro}. Finally, we summarise and conclude in Section \ref{sec:conc}. Readers interested primarily in the results, and not the technical details of the simulation and morphological metrics, may wish to skip directly to Section \ref{sec:macro}.

%% file: Simulation.tex
\section{The EAGLE Simulations}
\label{sec:sims}
\vspace{2ex}

Here, we describe briefly some of the most pertinent aspects of the EAGLE simulations for this study. A comprehensive description of the simulations is given in \citet[][S15]{Schaye15}\comment{ and \citet[][C15]{Crain15}}. Key properties of the runs used in this work are listed in Table \ref{tab:sims}. Unless otherwise stated, we present analysis of the fiducial EAGLE simulation in this work, hereafter Ref-100. This simulates the evolution of a cubic volume of side length 100~cMpc using the reference EAGLE physics model.

The EAGLE simulation suite uses a modified version of the {\sc Gadget-3} TreeSPH code (an update to {\sc Gadget-2}, \citealt{Springel05b}) to follow the co-evolution of gas and dark matter within periodic cubic volumes, varying resolution and parameters for star formation and feedback. Models are used to handle the unresolved aspects of heating and cooling of gas \citep{Wiersma09a}, star formation and stellar mass loss \citep{Schaye08, Wiersma09b}, and stochastic thermal feedback associated with stellar populations and AGN \citep{DallaVecchia12}. The model parameters are calibrated so as to reproduce the galaxy stellar mass function, galaxy sizes and the relation between galaxy stellar and black hole mass at $z=0$ (S15). The details of this calibration can be found in \citet{Crain15}. 

To mitigate the effects of numerical fragmentation, the Jeans scale is required to be at least marginally resolved (ie. above the smoothing scales of Table \ref{tab:sims}). In EAGLE this is achieved by imposing a polytropic equation of state for high density gas, $P_{\rm EoS} \propto \rho^{4/3}$, setting a minimum pressure for gas in the ISM. This effectively limits ISM gas temperature to $T \gtrsim 10^{4}$~K. As a result, the cool temperatures and 200-300~pc scale heights of molecular gas discs \citep[e.g.][]{vanderKruit11} are unresolved in EAGLE. For this reason we make no distinction between the vertical structure of discs, and define discs kinematically as the \textit{prograde excess} (section~\ref{sec:kins}).  

Halos are defined in EAGLE using the friends-of-friends (FoF) algorithm\comment{ \citep{Davis85}}, with substructures  (subhalos) identified using the SUBFIND algorithm \citep{Springel01, Dolag09}. We consider individual galaxies to be exclusive to individual subhalos and to comprise material within a 30~pkpc sphere about the galaxy centre, as detailed below.

\begin{table}
\begin{center}
\caption{Parameters of the EAGLE simulations used in this work. From left to right: simulation identifier, side length of cubic volume $L$ in co-moving Mpc (cMpc), gas particle initial mass $m_{\rm g}$, Plummer equivalent gravitational softening $\epsilon_{\rm prop}$ at redshift $z=0$ in  proper kpc (pkpc), and the paper reference for each volume.}
\label{tab:sims}
\begin{tabular}{lrrrr}
\hline
Name & $L$ & $m_{\rm g}$ & $\epsilon_{\rm prop}$& Ref. \\
& cMpc & $10^5 {\rm M}_\odot$ & pkpc & \\
\hline
RefL025N0376 (Ref-25) &  25 & $18.1$ & 0.70 & S15\\
RefL025N0752 (RefHi-25) &  25 & $2.26$ & 0.35 & S15\\
RecalL025N0752 (Recal-25) &  25 & $2.26$ & 0.35 & S15\\
RefL100N1504 (Ref-100) & 100 & $18.1$ &0.70 & S15\\
\hline
\end{tabular}
\end{center}
\end{table}

\subsection{Centring}
\label{sec:center}

An important aspect of the measurement of morphology is how the galaxy centre is defined. There are numerous ways in which this could be done, depending on the application. In this study we focus on both stars and gas, so we use both types of baryonic particles to define a common galaxy centre. In the majority of cases this value is not significantly different from using stars alone, particularly at lower redshifts. 

We utilise a shrinking spheres approach to define the galaxy centre, as in \citet{Trayford17}. This is an iterative process where once the baryonic centre of mass is found within a spherical aperture, an incrementally smaller aperture is re-centred and the centre of mass is re-defined. This starts with a large initial aperture of 100~pkpc and shrinks by $17\%$  each iteration until fewer than 200 baryonic particles are enclosed. We find that this procedure typically locates the mode of the mass distribution, analogous to using the brightest pixel in imaging data and refer to this as the \textit{`galaxy centre'} in what follows. The mass-weighted average velocity of the final sphere is also taken to be the galaxy peculiar velocity, and is subtracted from particles to define the rest-frame velocities when dynamical properties are being derived.

\subsection{Galaxy Sample}

To facilitate a meaningful study of morphology with the EAGLE simulations, it is important to select galaxies for which morphological properties are sufficiently resolved. A standard resolution criterion for galaxies in hydrodynamical simulations is that the number of resolution elements they comprise exceeds a certain threshold. However, as morphology pertains to the mass distribution in galaxies, the spatial resolution should also be considered. For example, compact galaxies may be more affected by gravitational smoothing than relatively extended galaxies of the same mass, as their mass profiles are better resolved spatially. 

 We focus on galaxies with stellar masses that exceed $10^{9} \; {\rm M_{\odot}}$, corresponding to $\gtrsim 500$ star particles per galaxy. We do not impose a limit on the compactness of galaxies because this would bias the morphological properties for a given mass, but rather investigate the convergence of morphological properties with both galaxy size and stellar mass in Appendix~\ref{ap:conv}. We find that the effect of spurious morphological properties in compact galaxies is small for the overall galaxy sample used here.

%% file: Morphologies.tex
\section{Characterising Galaxy Morphology}
\label{sec:morphs}

Quantifying the morphology of simulated galaxies and measuring the structures that comprise them requires metrics of morphology to be defined.  This itself presents some difficulty, owing to the multi-faceted and heterogeneous nature of galaxy morphology. Here we use morphology exclusively to refer to the \textit{physical} structure of galaxies; either the 3D distributions of mass in galaxies or the kinematic properties of galaxies that manifest them. In this study we first focus on measuring how \textit{`discy'} galaxies are. We consider metrics that probe this via \textit{kinematics}, quantifying the amount of material undergoing coherent and ordered rotation about the galaxy centre, or via the galaxy \textit{shape}, where the 3D stellar distribution is quantified. Again, this is investigated in the physical domain using the direct simulation output, as opposed to post-processed mock observations. We also identify disturbed morphologies that fall outside the HS classification using galaxy asymmetry.

\subsection{Kinematic metrics}
\label{sec:kins}

\begin{figure*}
	\includegraphics[width=1\textwidth]{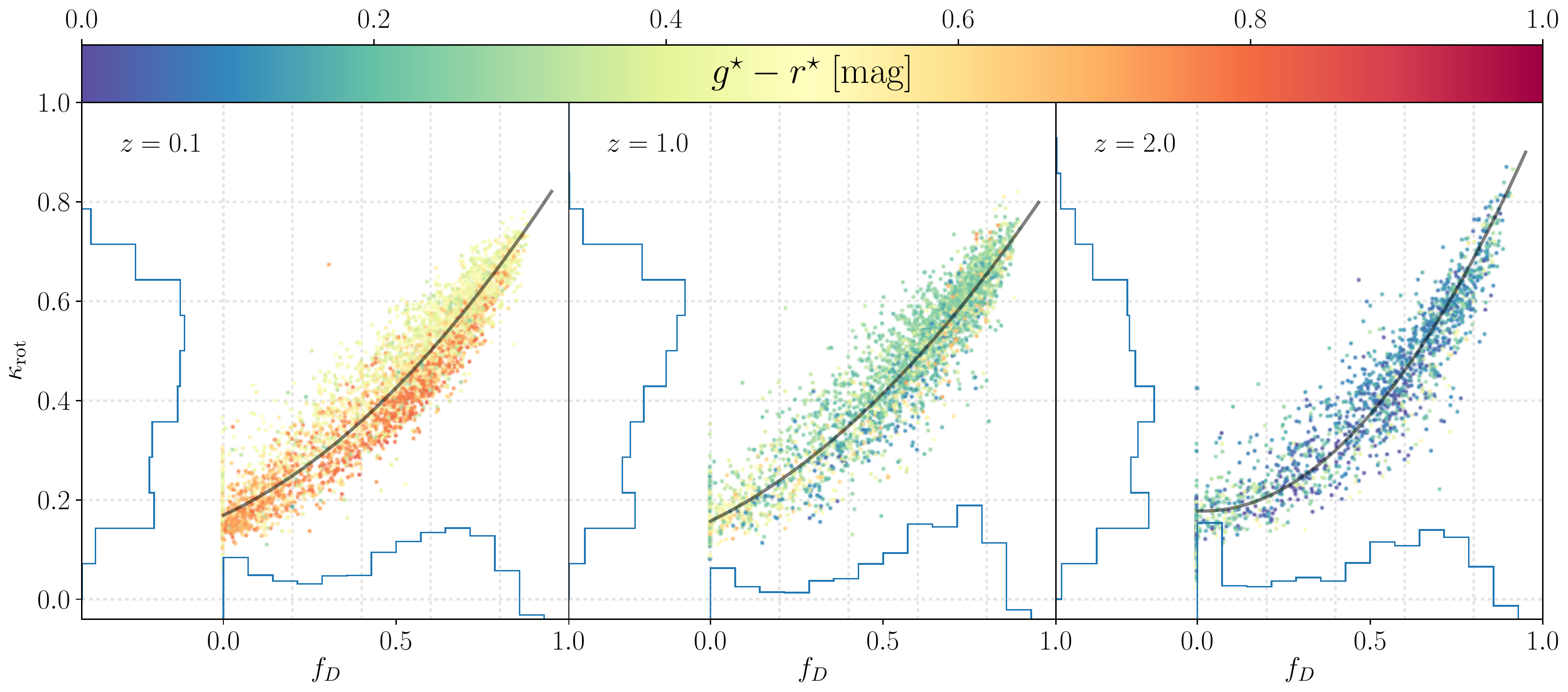}
\caption{The stellar disc-to-total ($f_D$) and $\kappa_{\rm rot}$ morphological metrics (see Section~\ref{sec:kins}) for the Ref-100 EAGLE galaxy sample in the stellar mass range $10 \leq \log_{10}(M^\star / {\rm M_\odot}) < 11 $ at $z=0.1$ (left), 1 (middle) and 2 (right). \textit{Points} indicate individual galaxies, with the projected $f_D$ and $\kappa_{\rm rot}$ histograms plotted on the $x$ and $y$ axes respectively. The measures correlate, and each implies the presence of both disc- and spheroid-dominated systems at every redshift. Galaxies are coloured by intrinsic $g-r$ colour, computed as described by \citet{Trayford15}. The distributions show little redshift evolution particularly for $f_D$.}
    \label{fig:comp}
\end{figure*}

A number of kinematic metrics have been devised to measure the ordered, or disc-like, rotation in simulated galaxies. A simple metric is the $\kappa_{\rm rot}$ parameter \citep{Sales10}, essentially a measure of the fraction of the kinetic energy in a galaxy that is in ordered rotation. However, \citet{Correa17} point out that this metric can be improved by not excluding material on retrograde orbits from this fraction. We use their improved prescription, where the rotating contribution is calculated only for prograde rotation. We measure this for all selected galaxies using the stellar material within a 30~pkpc spherical aperture about the galaxy centre.

Another technique is to appeal to the \textit{circularity}, $\epsilon$, of material within simulated galaxies. This compares the angular momentum of material along the net rotation axis of a galaxy to that of a circular orbit with the same energy \citep{Abadi03}. The circularity distribution is found to be generally bimodal in late types, with a peak around $\epsilon=0$ representing the spheroidal component, and a peak around $\epsilon\approx 1$ representing the disc component. To decompose the bulge and disc components,  \citet{Abadi03} estimate the bulge mass by doubling the mass in star particles with negative circularities, $2M^\ast(\epsilon < 0)$, and assuming the residual stellar mass, or \textit{prograde excess}, resides in a disc. Other authors explicitly select the disc component by using a positive threshold circularity value, e.g. $\epsilon \geq 0.65$ \citep{Tissera12}. Here we emulate the \citet{Abadi03} method\footnote{Measuring the \textit{prograde excess} requires only the \textit{sense} of the rotation relative to the net angular momentum, so for efficiency we only compute this for the majority of redshifts, and leave analysis of \textit{circularities} ($\epsilon$) at certain redshifts to Section \ref{sec:micro}.} to obtain the bulge and disc masses of galaxies, and take the disc stellar mass fraction, $f_D$, as a measure of morphology. The effects of disc identification are explored further in Section \ref{sec:micro}.

The $\kappa_{\rm rot}$ and $f_{\rm D}$ values for EAGLE galaxies in the stellar mass range $10<\log_{10}(M^\star/{\rm M_\odot})<11$ are compared directly in Fig.~\ref{fig:comp}. Individual galaxies are plotted for the Ref-100 volume at each of the $z \in [0.1, 1, 2]$ snapshots, and the histograms for these metrics are projected onto their respective axes, normalised to have an integral of 1. The shading of data points conveys the intrinsic $g-r$ colours of galaxies, computed as described in \citet{Trayford15}. We see that, as expected, the two metrics generally correlate well at each of the redshifts presented here\footnote{We note that this level of correlation persists for stellar masses below the restricted range displayed here.}.

Comparing the histograms in Fig.~\ref{fig:comp} reveals  striking consistency in the distributions of $f_D$ over the $0.1 \lesssim z \lesssim 2$ range. This consistency appears in spite of evolution in the mass distribution of galaxies with redshift, evidenced by the fewer galaxies populating the plot at $z=2$ compared to $z=0.1$ (and plotted in Fig.~\ref{fig:mfs} below). The $\kappa_{\rm rot}$ measure shows some evolution between $z=2$ and $z=1$, but is similarly consistent at $z \lesssim 1$.

Despite this lack of evolution, it is interesting to note the residual trend with colour that can be seen in the $z=0.1$ relation, such that galaxies with a higher fraction of kinetic energy in co-rotation for a given disc mass fraction are bluer. The trend between intrinsic $g-r$ colour and the $\kappa_{\rm rot}$-$f_D$ residual is in fact stronger (with a rank correlation coefficient of 0.46) than the trend  between colour and either metric individually (0.25 and 0.36 for $f_D$ and $\kappa_{\rm rot}$ respectively). This can be understood intuitively if galaxies with dynamically colder discs are more efficient at forming stars, and thus appear bluer on average. We explore the role of dynamically cold discs further in Section \ref{sec:micro}. Intriguingly, this trend is much weaker in the $z=1$ and 2 panels.

Given the general correlation between $\kappa_{\rm rot}$ and $f_{\rm D}$, we drop the $\kappa_{\rm rot}$ values and use $f_{\rm D}$ to represent a kinematic measure of morphology in the remainder of this work\footnote{An additional $f_D$ value is also computed for each galaxy using the \textit{zero-age} stellar mass (i.e. the zero-age main sequence mass of stars) to define the prograde excess, in order to compute the mass transfer rates between structures in Section \ref{sec:trans}.}. The $f_D$ histograms in particular show little evidence for strong evolution over the $0.1 \lesssim z \lesssim 2$ range, maintaining a relatively uniform distribution of galaxies with $0 \lessapprox f_D \lessapprox 0.9$ at each snapshot. Assuming the $f_D$ values correspond to the position of a galaxy along the HS thus suggests a consistent and diverse morphological mix at these stellar masses, from $t_{\rm lookback} \sim$~10~Gyr to the present, where early and late type galaxies exist with comparable number densities. This agrees with \citet{Clauwens17}, who calculate complementary bulge fractions ($1-f_D$) for EAGLE galaxies. While this paints a picture of a static HS that is established early on in the Universe, it is important to recognise that $f_D$ alone represents a restricted view of morphology: a dichotomy of kinematic discs and spheroids. To further explore the divergence from the Hubble sequence and this simple picture of morphology, we now turn to measures of galaxy shape and asymmetry. 

\subsection{Shape Metrics}

\begin{figure*}
\includegraphics[width=0.99\textwidth]{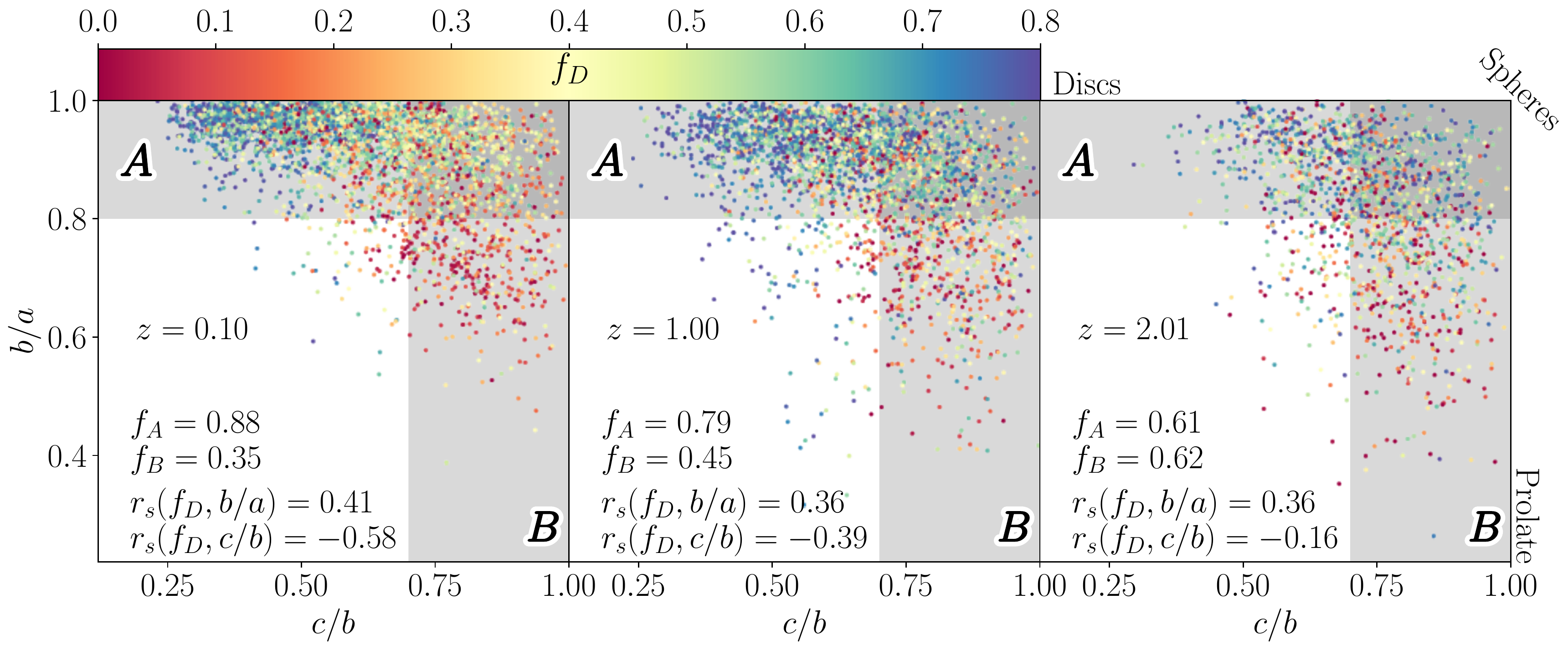}
\caption{The distribution of $10^{10}{\rm M_\odot} < M^\star < 10^{11}{\rm M_\odot}$  galaxies in terms of their axial ratios. Individual galaxies are coloured by their $f_D$ value, as shown in Fig.~\ref{fig:comp}, where blue indicates a disc fraction of $\geq 0.8$ and red indicates a disc fraction of 0. Galaxies are plotted at $z=2$ (right), 1 (middle) and 0.1 (left). Two shaded regions are defined using ad-hoc axial ratio values of $b/a > 0.8$ (region A) and $c/b > 0.7$ (region B). These divide more and less prolate galaxies (bottom and top, respectively) and more and less oblate galaxies (left and right, respectively). We see that at high redshift a higher proportion of galaxies are outside of region A (more prolate), and there is less correlation with kinematic morphology than seen at low redshift. For each redshift, the fraction of galaxies in each region and the rank correlation coefficient between each axis and $f_D$  are inset.}
   \label{fig:shapes}
\end{figure*}

In addition to kinematic measures, the shape of a galaxy may be used as a more direct measure  
of morphology. Exploiting the complex 3D mass profiles of the galaxies that emerge in the EAGLE simulations, we can calculate the moment of inertia tensor:
\begin{equation}
\hat{\boldmath I} = 
\begin{bmatrix}
    \sum\limits_{i=1}^N (y^2_i + z^2_i)m_i        
    & \sum\limits_{i=1}^N x_i y_i m_i 
    & \sum\limits_{i=1}^N x_i z_i m_i   \\
    \sum\limits_{i=1}^N y_i x_i m_i       
    & \sum\limits_{i=1}^N (x^2_i + z^2_i) m_i 
    & \sum\limits_{i=1}^N y_i z_i m_i   \\
    \sum\limits_{i=1}^N  z_i x_i m_i       
    & \sum\limits_{i=1}^N z_i y_i m_i 
    & \sum\limits_{i=1}^N (x^2_i + y^2_i) m_i \\
\end{bmatrix} ,
\end{equation}
where $m_i$ represents the mass of the $i$th star particle, and $(x_i, y_i,z_i)$ are the particle coordinates measured from the galaxy centre (calculated as in Section~\ref{sec:center}). Each tensor term is summed over the $N$ star particles that comprise the galaxy.

The eigenvectors and eigenvalues of the diagonalised inertia tensor yield the primary axes of the galaxies and their associated moments of inertia, respectively. These moments of inertia characterise the mass distribution perpendicular to each axis.  The primary moments of inertia are referred to as $I_1$, $I_2$ and $I_3$ where $I_1\geq I_2\geq I_3$. Two axial ratios can then be expressed as 
\begin{equation}
b/a = \sqrt{\frac{I_1+I_3-I_2}{I_2+I_3-I_1}}, \\
c/b = \sqrt{\frac{I_1+I_2-I_3}{I_1+I_3-I_2}},
\end{equation}
where $a$, $b$ and $c$ ($a\geq b\geq c$) represent the lengths of the primary axes. This triaxial approach provides a simplified but descriptive parametrisation of the 3D shape. 
\begin{figure*}
\includegraphics[width=0.99\textwidth]{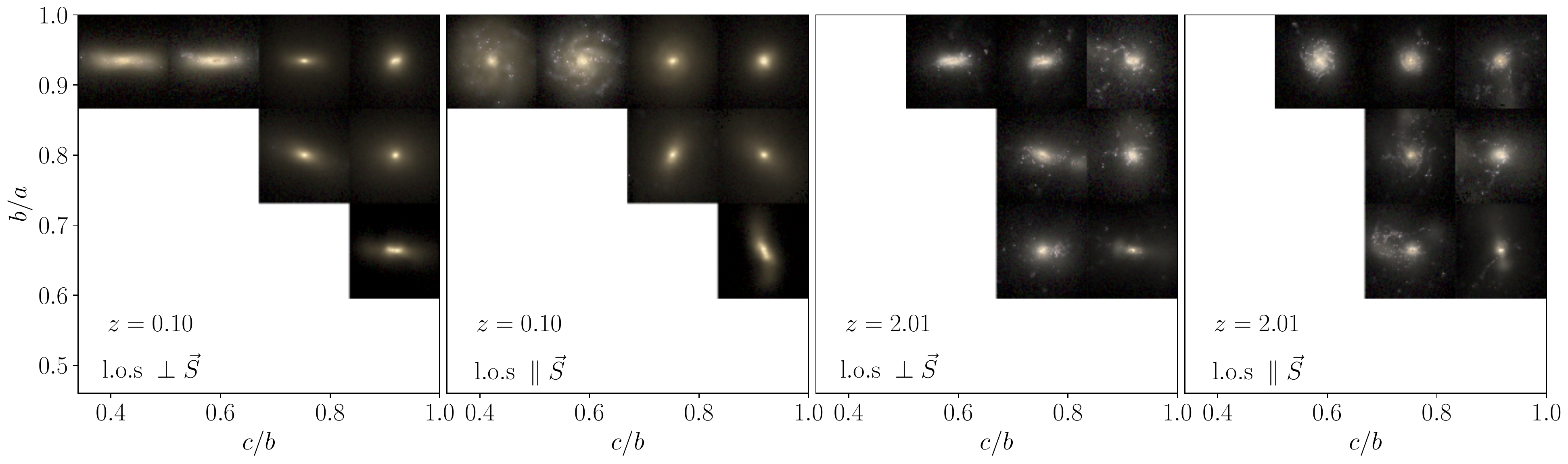}\vspace{2ex}
\includegraphics[width=0.99\textwidth]{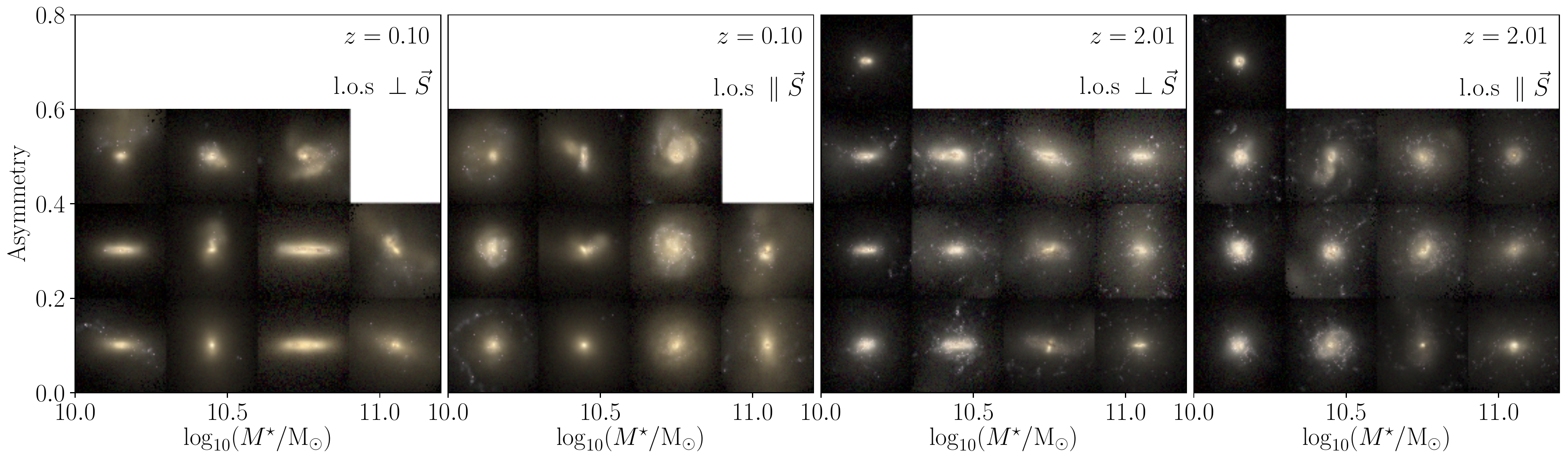}
    \caption{Visualising galaxies with differing shape properties. The top row shows the axial ratios of Fig.~\ref{fig:shapes}, using $gri$ composite images to visualise the median-sized galaxy contained within the region covered by each image. Each image shows a $60\times 60$~pkpc field of view about the galaxy centre. The two left and two right panels indicate galaxies at $z=0.1$ and 2 respectively, and show the same galaxies projected perpendicular and parallel to the galaxy spin vector $\vec{S}$ in each panel. The bottom row uses the same format, except now galaxies are distributed in the plane of asymmetry (Eq.~\ref{eq:asym}) and $\log_{10}(M^\star / M_{\odot})$. The top row shows galaxies exhibiting different shape parameters, demonstrating that the prolate and triaxial systems possess a mixture of disturbed and smooth profiles. The bottom row shows how these can be separated, with disturbed systems showing higher asymmetry.}
    \label{fig:imgrid}
\end{figure*}

We plot these axial ratios for EAGLE galaxies in Fig.~\ref{fig:shapes}, yielding a plane of galaxy shapes. Plotting $b/a$ as a function of $c/b$ yields a plane spanning the range $[0,1]$ along both axes. In this plane, galaxies approaching the top right are close to spherical, top left are more oblate (late-type or disc shaped), bottom right are more prolate and bottom left are triaxial. EAGLE galaxies are plotted for the same $z \in [0.1, 1, 2]$ redshifts and $10 \leq \log_{10}(M^\star / {\rm M_\odot}) < 11 $ range of Fig.~\ref{fig:comp}, and coloured by their $f_D$ value.  

First inspecting the $z=0.1$ (leftmost) panel, we see that most galaxies fall in the upper half of the allowable range in both axes ($> 0.5$). Galaxies with lower $c/b$ ($c/b\lessapprox$0.7), typically have high $b/a$ ($b/a\approx$0.95), indicative of oblate shapes. Conversely, higher $c/b$ galaxies ($c/b\approx$0.9) exhibit a broader range of $b/a$ values, indicating galaxies that range from near spherical to highly prolate. We see that the $f_D$ values correlate well with the $c/b$ axis in particular, such that galaxies with discy shapes have higher $f_D$ while spherical and particularly prolate galaxies exhibit lower $f_D$.

To demonstrate how the distribution of galaxy shapes evolves, we define two axial ratio cuts. A cut at $b/a=0.8$ separates those that are less prolate (region A) from those that are more prolate, and a cut at $c/b=0.7$ separates those that are less oblate (region B) from those that are more oblate. These separation values are somewhat \textit{ad-hoc}, but serve to show how the distribution in this plane and its correlation with $f_D$ evolves.     

The distribution of galaxies in the $z=1$ and $z=2$ panels reveals an evolution in the proportion of less prolate (region A) galaxies over cosmic time, from 61\%{} at $z=2$ to 88\%{} at $z=0.1$. There also appears to be a change in how well the $f_D$ values correlate with galaxy shape. By calculating the Spearman rank correlation coefficients between $f_D$ and both $c/b$ and $b/a$ at each redshift (inset in Fig.~\ref{fig:shapes}),  we see that these correlations become weaker with increasing redshift, and that the stronger correlation changes from $b/a$ at $z=2$ to $c/b$ at $z=0.1$. 

Considering the shapes of EAGLE galaxies in this way challenges the picture painted by Fig.~\ref{fig:comp} of a static morphological mix at $z \leq 2$. We see that while the $f_D$ distribution remains relatively constant, the shape distribution of galaxies evolves significantly, with $f_D$ becoming better correlated with galaxy shape as time progresses. 

In Fig.~\ref{fig:imgrid}, we use mock galaxy images for further insight into this nuanced picture of morphological evolution. These are rest-frame $gri$ images including dust radiative transfer as described in \citet{Trayford17}. The top row uses the same axes as Fig.~\ref{fig:shapes}, now showing an image of an example galaxy to represent each $b/a$-$c/b$ bin containing $> 3$ galaxies. The example galaxy is selected to be the galaxy with the median galaxy size\footnote{Using the 3D half-mass radius from \citet{Furlong17}.}. This is preferred to randomly selected galaxies as it yields a deterministic selection which minimises stochastic size variations between bins. The left and right pairs of plots show $z=0.1$ and $z=2$ galaxies respectively, with the respective leftmost and rightmost panels in each pair displaying the same galaxies viewed along lines of sight parallel and perpendicular to the net spin vector of the stars. 

The top row in the low-redshift ($z=0.1$) sample resembles familiar HS morphologies, from disc-like late-types to near spherical early-types. The more prolate galaxies appear as increasingly elliptical early types, with smooth light distributions. 

The higher-redshift galaxies show significant visual differences. Galaxies appear typically smaller at $z=2$, particularly comparing the late-type galaxies in the top two rows to those at $z=0.1$. There appears to be a less clear difference in the appearance of oblate and spherical galaxies along the top row, which may also be an effect of the smaller sizes combined with resolution limitations. A subtle difference between the leftmost and rightmost galaxies in the top row is the presence of a faint stellar halo in the spherical galaxy when viewed perpendicular to the spin vector. The high-redshift galaxies also seem to show generally clumpier profiles. This could in part be due to the greater prevalence of bright star-forming regions, but the clumpier diffuse light indicates that this is also down to these galaxies being typically more disturbed. It is important to note that while dust has no bearing on the intrinsic stellar shapes we measure, it can affect the visual morphologies in these images.

Perhaps the most striking difference is between the prolate galaxies ($b/a < 0.8$) at $z=0.1$ and $z=2$. At high redshift the prolate shapes appear driven by large-scale disturbances, unlike the smooth prolate galaxies seen at low redshift that may be considered a variety of spheroid. However, it is clear that a disc-spheroid decomposition of disturbed galaxies has little physical meaning, and may produce misleading results, particularly at high redshift where disturbed galaxies are more prevalent. Instead, these systems should be considered a separate morphological category. Since the triaxial approach cannot effectively separate these systems at all redshifts, we develop a more direct measure of disturbance using galaxy asymmetry next. 

\subsubsection{Shape Asymmetry}

The degree of asymmetry in galaxy light profiles and its connection to galaxy colour and interaction history has been explored observationally \citep[e.g.][]{Conselice00, Pawlik16}. \citet[][]{Conselice00} define asymmetry by subtracting a CCD image of a galaxy from a copy rotated by 180$^\circ$, and measuring the ratio of absolute residual flux to the total object flux. This functions as a measure of the point symmetry of galaxy pixels about the image centre. For our purposes we develop a 3D analog to the asymmetry parameter. This is more consistent with our 3D measurement of stellar kinematics and shapes, keeping the measurement independent of viewing angle and projection effects such as dust obscuration.

Using the {\sc healpix} code \citep{Gorski05} to define uniform bins of solid angle about the galaxy centre, we calculate the total stellar mass in each bin. The asymmetry is then computed by summing the absolute mass difference between diametrically opposed bins, and dividing by the total stellar mass. Using the RING numbering system for {\sc healpix} bins \citep{Gorski99}, this can be written as,

\begin{equation}
A_{\rm 3D} = \frac{\sum_{i = 1}^{N/2}  \lvert m^\star_i-m^\star_{N-i} \rvert}{\sum_{i=1}^N m^\star_i},
\label{eq:asym}
\end{equation}
summing over all bins, where $m^\star$ represents the stellar mass in each bin.  This could be adopted for any {\sc healpix} resolution level, but here we choose the minimum 12 bins.

Centring is particularly important for the measure of $A_{\rm 3D}$. Centring using the centre of mass of all star particles would, by definition, minimise the asymmetry parameter, so our use of a `shrinking-centroid' approach is preferable. This will be closer to the \textit{mode} of the stellar density distribution. It is important to note that this could still give spurious results in some cases, for example by centring the galaxy on a very dense clump of particles offset from the `true' galaxy centre.  

While it is also a measure of point symmetry, this $A_{\rm 3D}$ parameter is not directly analagous to the 2D  metric of \citet[][]{Conselice00}, as there is no radial binning of material within each solid angle bin. The use of only 12 bins also means that this measure is coarse relative to the number of pixels often used to calculate the 2D asymmetry. The choice to coarsely bin the stellar material is made to reduce sensitivity to small-scale angular variations, which are more prone to resolution effects. Nevertheless, we tested this parameter and found that it recovers the asymmetry in idealised test cases where a spheroidal particle distribution is perturbed by adding a lower-mass off-centre spheroidal particle distribution to it. The random uncertainty in $A_{\rm 3D}$ depends on the number of particles as $1/\sqrt{N}$, so is $\approx 0.05$ for an EAGLE galaxy of $10^{9} {\rm M_{\odot}}$ at fiducial resolution.

The bottom row of Fig.~\ref{fig:imgrid} is produced in the same way as the top row, except now for the distribution of $A_{\rm 3D}$ vs. $\log_{10}(M^\star / {\rm M_\odot})$. We see that typically galaxies with $A_{\rm 3D} \gtrapprox 0.2$  show disturbed visual morphologies, indicative of recent or ongoing mergers. We also see how $A_{\rm 3D}$ can be used to help separate the spheroidal and disturbed prolate galaxies; the 2$^{\rm nd}$ and 4$^{\rm th}$ galaxies from the left in the lowest $A_{\rm 3D}$ bin in the $z=0.1$ sample show clear prolate spheroids.

\begin{figure}
\includegraphics[width=0.49\textwidth]{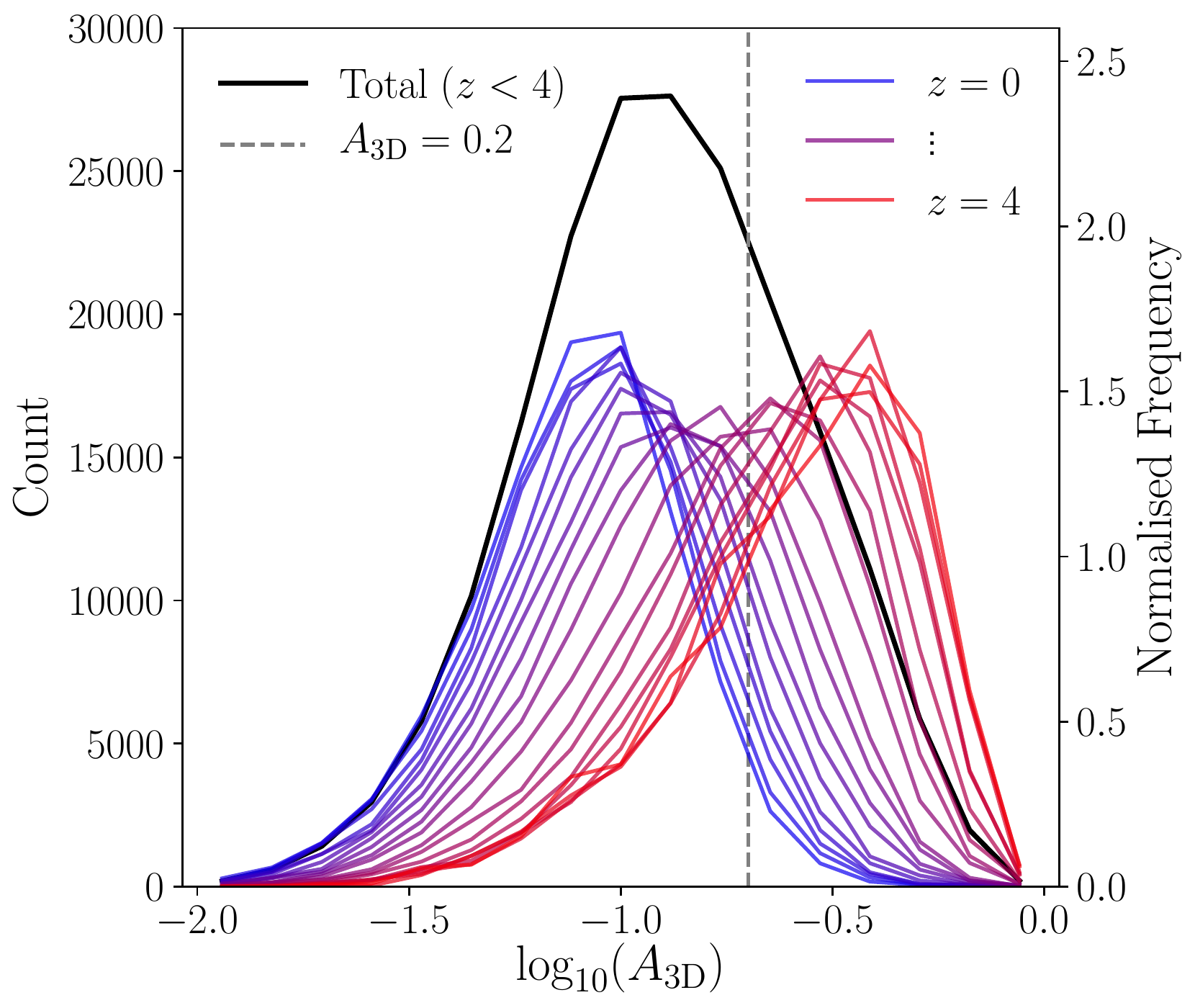}
    \caption{Histograms of $\log_{10}(A_{\rm 3D})$, Eq.~\ref{eq:asym}, for galaxies with $\log_{10}(M^\star/{\rm M_\odot}) > 9$. The black line shows the counts in each bin (left axis) for all galaxies at $z < 4$. Coloured lines show separate histograms for each simulation snapshot (normalised to unit integral, right axis). Our threshold of $A_{\rm 3D} = 0.2$ is indicated by the dashed line.}
\label{fig:a3ds}
\end{figure}

We can select disturbed galaxies using a cut in $A_{\rm 3D}$. We plot $\log_{10}(A_{\rm 3D})$ distributions in Fig.~\ref{fig:a3ds}. While the overall histogram of galaxies at $z<4$  is unimodal, we see evolution in the peak position, with $A_{\rm 3D} \approx 0.1$ at low redshift ($z \lesssim 0.6$) and $A_{\rm 3D} \approx 0.3$ at high redshift ($z \gtrsim 2$); a value of $A_{\rm 3D} \approx 0.2$  is chosen to divide these regimes.

Galaxies with $A_{\rm 3D} > 0.2$ are hereafter referred to as \textit{asymmetric} systems. All other galaxies are taken to be \textit{Hubble sequence} (HS) members, for which a disc/spheroid decomposition is deemed appropriate. While this choice is somewhat \textit{ad-hoc}, we find that different cuts yield qualitatively similar results, as we explore further in appendix \ref{ap:pecs}. These high-$A_{\rm 3D}$ galaxies are also  a potentially useful sample for future studies of disturbed morphologies and merger remnants. 

%% file: MacroMorph.tex
\begin{figure*}
	\includegraphics[width=1\textwidth]{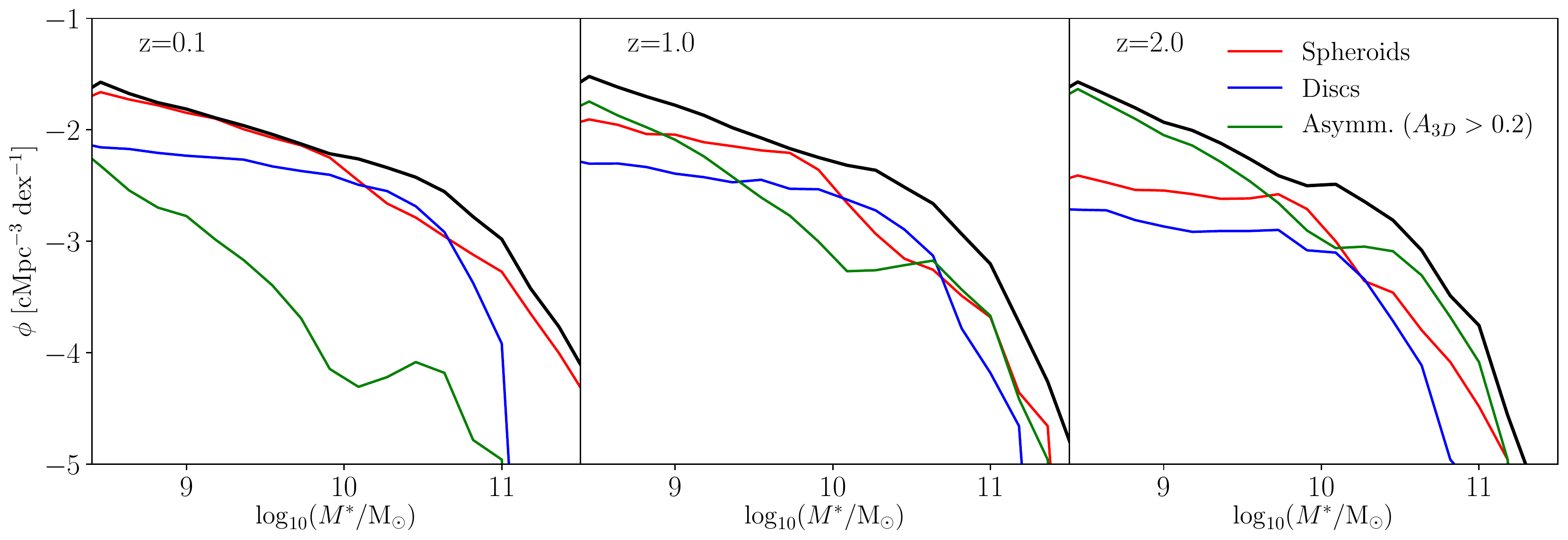}
    \caption{The evolution of the galaxy stellar mass function (black lines) alongside the mass function of different morphological components.  Disc structures (blue lines) and spheroid structures (red lines) are separated using the \textit{prograde excess}, and each Hubble sequence galaxy contributes separately to both mass functions,  whereas \textit{asymmetric} galaxies (green lines) contribute their whole mass\inorig{in compact and extended classes}. As redshift increases the decreasing contribution of Hubble sequence galaxies, and the increasing contribution of asymmetric systems, can be seen. \inorig{The split in asymmetric types effectively separates high and low mass asymmetric types.}}
    \label{fig:mfs}
\end{figure*}

\begin{figure}
\includegraphics[width=0.48\textwidth]{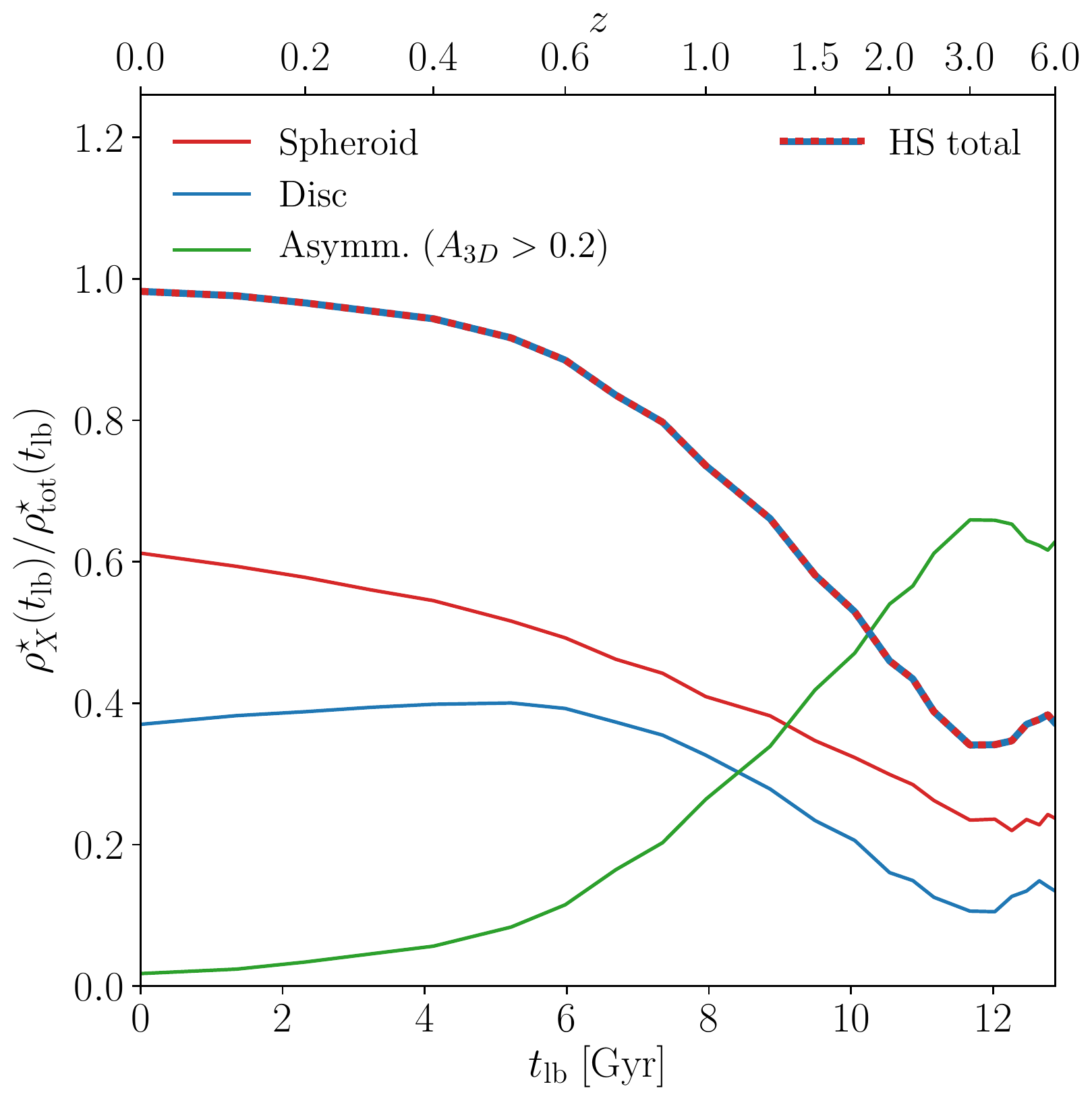}
    \caption{The evolving fractional contribution of structures to the overall cosmic stellar mass density, measured for galaxies of $M^\star > 10^{9}{\rm M_\odot}$, with the contributions of spheroids and discs in Hubble sequence galaxies shown in red and blue respectively. For reference, the total Hubble sequence (red-blue dotted) and asymmetric (green) contributions are also displayed. The transition to Hubble sequence dominance can be seen around $z=1.5$, with asymmetric types comprising $\sim 5\%$ of the mass at the present day.}
    \label{fig:cum}
\end{figure}

\begin{figure*}
	\includegraphics[width=1\textwidth]{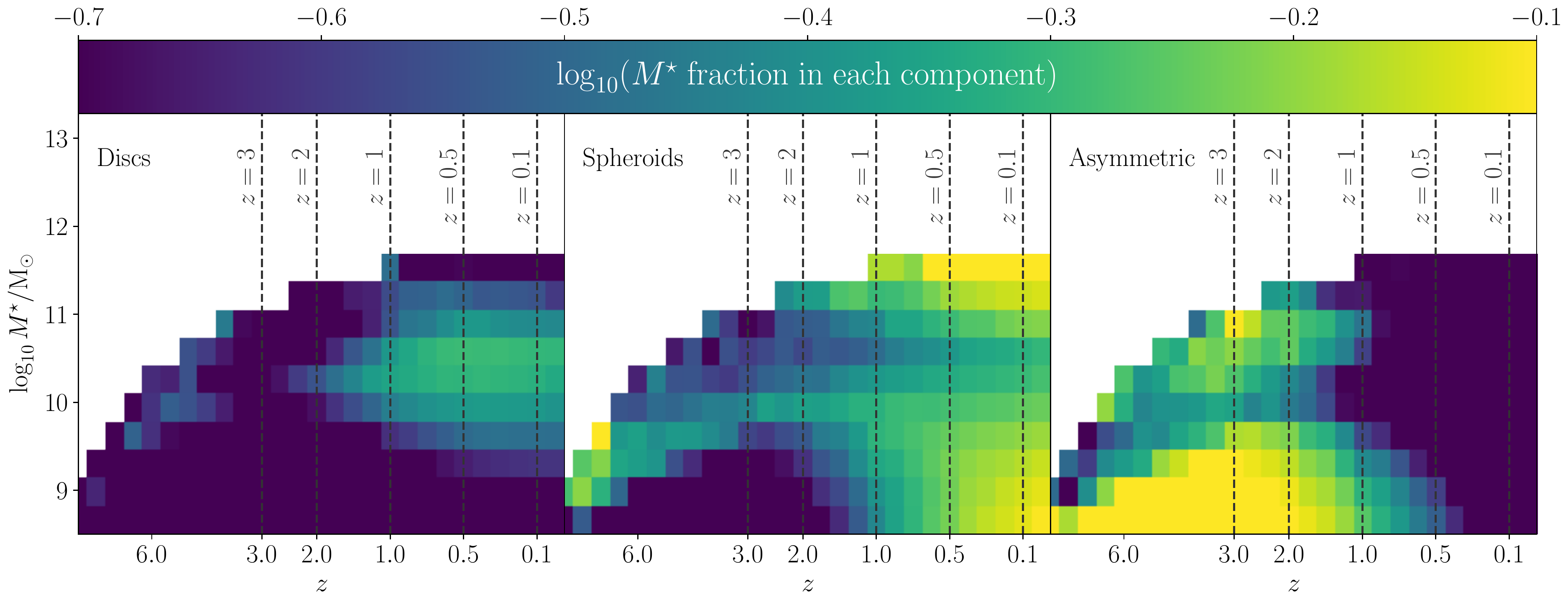}
    \caption{The evolving fraction of stellar mass in particular morphological structures (different panels) as a function of galaxy stellar mass and redshift. Panels from left to right show the logarithmic fraction of stellar mass in disc, spheroid and asymmetric structures. Separating asymmetric galaxies from the conventional disc and spheroid structures that comprise Hubble sequence galaxies reveals how the HS emerges at $z\approx2$. The fraction in discs peaks at a relatively constant mass of $M_\star \approx 10^{10.3} {\rm M_{\odot}}$.}
    \label{fig:frac}
\end{figure*}

\section{The evolving mass content in different morphological structures}
\label{sec:macro}
In the previous section we considered various metrics for morphology, and gained insight into how these measures represent the morphological evolution of galaxies.  We now turn to a more detailed look at the evolution of morphological \textit{structures} through cosmic time.

Fig.~\ref{fig:mfs} shows the stellar mass function of \textit{disc}, \textit{spheroid} and \textit{asymmetric} morphological structures at redshifts 0.1, 1 and 2. These are plotted alongside the overall galaxy stellar mass function. To obtain these, we first define HS galaxies to be below a threshold asymmetry value of $A_{\rm 3D} \leq 0.2$ (see section~\ref{sec:morphs}). The mass in HS galaxies is then divided into bulge and disc components using the disc fraction\footnote{Measured as the \textit{prograde excess}, or the total mass minus twice the counterrotating mass, see section~\ref{sec:morphs}.}, $f_D$, with each component contributing separately for each galaxy. It is important to emphasise that these mass functions are distinct from the mass functions of whole galaxies separated by some $f_D$ cut. For asymmetric galaxies ($A_{\rm 3D} > 0.2$), the whole galaxy contributes to the mass function. To complement this plot, we also show the evolving cosmic stellar mass density in each of these structures in Fig.~\ref{fig:cum}, computed by integrating these mass functions at each snapshot. We plot the mass density in HS galaxies (red-blue line) and asymmetric galaxies (green-orange line) in addition to each of the four morphological structures and the total evolution.

At $z=0.1$, we see that \textit{disc} and \textit{spheroid} components of HS galaxies dominate the total mass function over the whole mass range. Galaxy bulges are more prevalent than discs at all mass, except at around $\log_{10}(M^\star/{\rm M_\odot}) = 10.3$ where discs are the most common structure. Comparing to Fig.~\ref{fig:cum} at $z=0.1$ (leftmost points), we see that \textit{spheroids} are the dominant morphological component with discs holding $\approx40$~\%{} less stellar mass. We also see that the asymmetric galaxies contribute most at the low-mass end but they make up only 5\% of the mass. We note that the absolute contribution of asymmetric galaxies depends on the value of the $A_{\rm 3D}$ threshold that we have used to define them. As such, it is pertinent to instead focus on how the relative contribution of asymmetric galaxies varies with mass and redshift. The $z=1$ and $z=2$ panels of Fig.~\ref{fig:mfs} show how the normalisation of the mass function of \textit{asymmetric} galaxies evolves strongly, but retains a similar characteristic shape. 

The evolution in the mass contributions of different morphological structures is further visualised in Fig.~\ref{fig:frac}, which shows the stellar mass fraction contributed by separate structures in bins of \textit{galaxy}\footnote{As opposed to the stellar masses of the structures themselves that are used to construct the Fig.~\ref{fig:mfs} mass functions.} stellar mass for each output snapshot. We see from Fig.~\ref{fig:cum} that HS galaxies (red-blue line) come to dominate over asymmetric galaxies (green line) at $z\approx 1.5$. Fig \ref{fig:frac} shows that, as HS galaxies become prominent, discs come to dominate at a characteristic galaxy stellar mass of $\log_{10}(M^\star/{\rm M_\odot}) \sim 10.5$, as was also shown by \citet{Clauwens17}. The disc fraction continues to peak at this mass until the present day, while also contributing a growing fraction to lower mass bins. Fig.~\ref{fig:cum} shows that the overall fraction of stellar mass in discs plateaus at $z \approx 0.6$. The fraction of mass in spheroids grows steadily for $z \lesssim 3$, coming to dominate both the highest and lowest mass bins by $z=0$.

Taken together, Figs.~\ref{fig:mfs}, \ref{fig:cum} and \ref{fig:frac} paint a picture of a HS that rises to prominence at $z \approx 2$ and comes to dominate at low redshift ($z \lessapprox 1.5$), as seen explicitly in the cosmic stellar mass density contributions of Fig.~\ref{fig:cum}. While the exact $A_{\rm 3D}$ threshold for asymmetric galaxies is debatable, we see that the fraction of asymmetric systems exhibits significant trends with mass and redshift. Asymmetric galaxies dominate the stellar mass budget at early times in EAGLE, and contribute only marginally at $z \approx 0$. For HS galaxies, the fractional increase of cosmic stellar mass is almost monotonic with cosmic time. However, the fraction of this stellar mass that is in discs peaks at $z \approx 0.6$, and the total mass fraction  declines very slightly for $z \lessapprox 0.4$. At redshifts $z \lessapprox 0.6$  spheroidal structures constitute the majority of the cosmic stellar mass density. 

The ensemble evolution of morphologies described by these results supports a three-phase schematic model of galaxy formation in EAGLE similar to that of \citet{Clauwens17}, but where we see trends in redshift as well as stellar mass. The initial \textit{assembly} of galaxies and their halos dictates morphology at high redshifts, with most galaxies exhibiting asymmetric morphologies before dynamical relaxation process can act. At intermediate redshifts $z\approx1-2$, the formation of coherent gas discs from the higher angular material accreting onto the halos leads to rampant star formation and the emergence of HS late-type galaxies as the stellar mass in discs grows rapidly. As time progresses, falling gas fractions slow star formation in discs. Without continued replenishment, the ongoing decay of ordered stellar discs by successive mergers, galaxy interactions and secular processes lead to the decline of absolute stellar mass in discs at $z \lessapprox 0.3$. This disc destruction continues to feed the slowing growth of HS spheroids towards the present day, where the fraction of mass in spheroids is highest. 

In the following sections we will examine this picture of morphological evolution by considering the evolution of individual galaxies, as well as the statistics of morphological change, by comparing where stars are born to where they reside at later times.

%% file: TrackMorph.tex
\section{Exploring morphological change in individual galaxies}
\label{sec:tree}

The previous section described how the prominence of different morphological components evolves in the EAGLE simulation, supporting the three phase evolution found by \citet{Clauwens17}. In this paradigm, kinematically disordered low-mass galaxies grow to become disc-dominated systems forming stars in-situ, until reaching a mass where the star formation efficiency drops and galaxies become largely spheroidal. In order to understand the mechanisms driving this change, we first try to understand the nature of morphological transformation in individual galaxies.

\begin{figure*}
\includegraphics[width=0.8\textwidth]{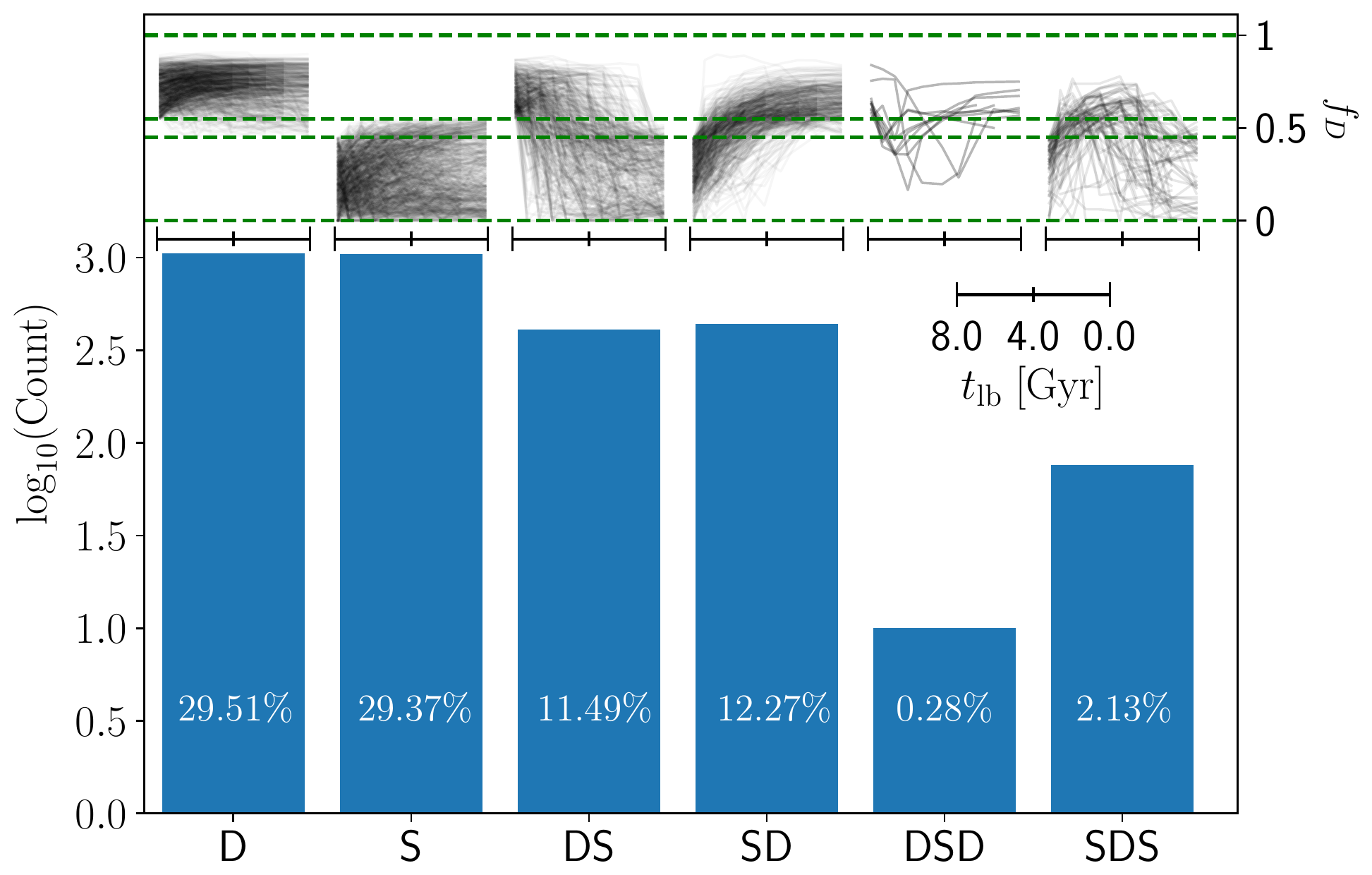}
    \caption{Classifying the morphological evolution of individual EAGLE galaxies between $z = 1$ and $z=0$, for galaxies with $M_\star > 10^{10}\; {\rm M_\odot}$ at $z=0$. This classification uses the $f_D$ ratio to measure transitions between disc-dominated (D) and spheroid-dominated (S) states. The sequence of letters labelling each bar represents the chronological sequence of states held by galaxies in that class at $z < 1$, e.g. D denotes galaxies that remain disc-dominated whereas DS galaxies transition from disc- to spheroid-dominated. Bar height denotes the $\log_{10}$ count of galaxies in each class. Galaxies contribute to only one class, based on their evolutionary histories. The tracks of individual galaxies in each class as a function of time are visualised by the translucent lines above each bar. Upper and lower horizontal dashed lines represent 0 and 1 respectively in $f_D$. Inner lines denote the transition region of 0.45 to 0.55 that must be traversed to register a transition. Tracks reveal common features between galaxies. In particular, the DS class shows rapid transitions but with a wide range of transition times, whereas SD shows a more gradual uniform trend. See text for discussion.}
    \label{fig:bars}
\end{figure*}

In Fig.~\ref{fig:bars} we attempt to visualise the morphological histories of EAGLE galaxies, by tracing the evolution of the $f_D$ parameter through cosmic time. Here, the progenitors of galaxies with $M_\star > 10^{10}\; {\rm M_\odot}$ at $z=0$ are traced back in cosmic time through each snapshot to $z=1$. As galaxy asymmetry is not considered in this plot, we stop at this redshift above which asymmetric galaxies begin to dominate the stellar mass budget (Fig.~\ref{fig:frac}). We use the evolution of the $f_D$ parameter to classify tracks of galaxies according to whether they are disc- or spheroid -dominated, with a change registered when a galaxy passes completely through the intermediate $0.45<f_D\leq 0.55$ region. Due to this requirement we exclude galaxies that are in the $0.45<f_D\leq 0.55$ range at $z=1$ ($\approx 10\%$). Galaxies that remain classified as spheroid dominated ($f_D < 0.55$) are labelled S; persistent discs ($f_D > 0.45$) are labelled D. Galaxies that transition between these states are labelled by a letter sequence reflecting their sequential change in status from $z=1$ (e.g. SDS). Bars denote the number of galaxies from the Ref-100 volume in each evolutionary class, with the fraction in each class inset, while the individual tracks are visualised along the top of this plot. The $f_D$ threshold values of 0, 0.45, 0.55 and 1 are indicated from bottom to top using horizontal lines.

We see that over this period the number of galaxies without a change between disc- and spheroid-dominated states (D and S) constitute about 60\% of the population, and are a factor 2.5 more common than those registering a single change (DS and SD). Multiple changes are less common still, accounting for $\lessapprox 5\%$ of galaxies. Inspecting the plotted $f_D$ tracks for individual galaxies in each class is especially informative for the DS and SD transitions.  The galaxies transitioning from  spheroid to disc show a coherent and gradual change, whereas the DS galaxies generally show a more stochastic and rapid change. The general behaviour of the SD class can be attributed to the independent growth of galaxies into a disc-dominated phase from an earlier, kinematically disordered phase, as described in \citet{Clauwens17}.

Conversely, the DS galaxy behaviour suggests a triggered rapid morphological transition, occurring stochastically. Mergers or galaxy interactions are strong candidate mechanisms for triggering the transition. \citet{Clauwens17} found that statistically mergers contributed to morphological change mostly through the growth of the kinematic spheroid rather than the destruction of discs. We find that 40\% of DS transitions experience a 10:1 or greater merger since $z=1$, compared to 18\% for a mass-matched control sample. These mergers also typically have lower median gas fractions, with 7\% relative to 12\% measured for the control. This is consistent with the finding that gas-poor mergers in particular reduce the specific angular momentum of stars in EAGLE galaxies \citep{Lagos17b}. Despite significant mergers being more common in DS galaxies, 60\% of the DS class show more quiescent histories, suggesting that mergers are not solely responsible for these transitions.
  
While this analysis gives some clues towards the processes driving morphological change, we merely aim to characterise the way in which disc fractions evolve in individual galaxies and leave a detailed study of the transformation mechanisms to a future work. Instead, we focus more generally on whether morphological structures are built predominately through internal star formation, or through transformational processes such as mergers or secular evolution. These effects can be disentangled statistically by assessing the fraction of star formation that takes place in each structure relative to the fraction of the stellar mass they account for. We investigate this in the following section.

%% file: MicroMorph.tex
\begin{figure*}
\includegraphics[width=0.98\textwidth]{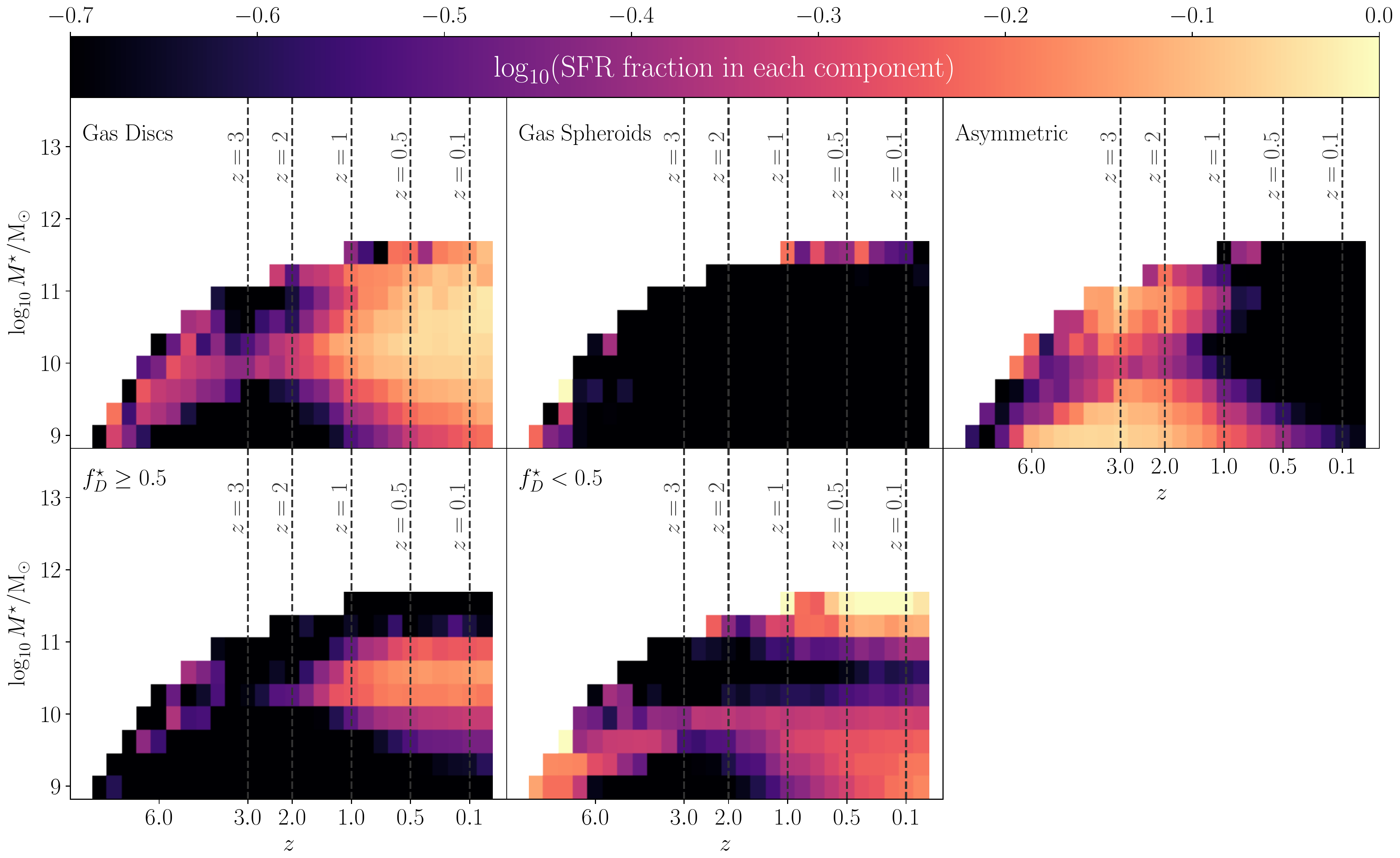}
    \caption{As Fig.~\ref{fig:frac}, except bins are shaded by the logarithmic fraction of the total star formation contributed by each labelled component. In the top row the contribution of gas disc and spheroid structures in HS galaxies are measured using the prograde excess star formation rate in gas particles, as opposed to the prograde excess stellar mass used in previous sections. In the bottom row the HS star formation is now split into contributions from `earlier'  type (left) and `later' type (right) galaxies via a stellar $f^\star_{\rm D}$ cut, for comparison. In contrast to Fig.~\ref{fig:frac}, the contribution of gas spheroids is minimal across all bins, except for some of the the highest mass bins. Asymmetric galaxies are found to contribute more to star formation in both the higher redshift and lower mass bins. However, the $z<2$ contribution of gas discs is striking at all $M^\star$.}
    \label{fig:sfrfrac}
\end{figure*}

\section{Stellar morphologies at birth and the role of transformational processes}
\label{sec:rates}

The fraction of the cosmic stellar mass budget residing in different morphological components at a given time is not necessarily a reflection of their contribution to ongoing star formation. 
Galactic discs are commonly seen as the sites of the majority of star formation at low redshift; however, discs host a subdominant fraction of the stellar mass in EAGLE, as shown in Fig.~\ref{fig:cum}. Other modes of star formation, such as starbursts in nuclear regions of galaxies and star formation in the tidal structures of merger remnant (or otherwise asymmetric) systems, may be non-negligible. In particular, the dominant role of star-forming gas discs may give way to less ordered morphologies at high redshift. This could help explain the dominance of asymmetric stellar systems at high redshift. In this section we explore the morphological properties of the star-forming gas through cosmic time, relative to that of the stellar component explored in the previous sections. 

\subsection{Star-forming gas morphologies}

The morphologies of the star-forming gas in EAGLE can be characterised using similar methods to those applied to the star particles,  outlined in Section \ref{sec:morphs}. The same threshold of $A_{\rm 3D} > 0.2$ (measured for stars) is used distinguish HS from asymmetric galaxies, such that the total star formation contributed by asymmetric galaxies can be easily determined. To separate the spheroid and disc contributions to the cosmic star formation rate density in HS galaxies, we again follow the ansatz of \citet{Abadi03} that the counter rotating material reflects half of the contents of the spheroidal component. However, we measure the fraction of the star formation rate (rather than the mass fraction) in counter-rotating gas particles, and double this to estimate the total fraction of the star formation rate associated with the spheroid\footnote{While this scheme could in theory yield nonsensical spheroid SFR fractions of $>1$, in practice this never occurs.}. The remainder of the star formation is then assumed to take place in the disc. 

The star formation in gaseous disc and spheroid structures calculated in this way is not equivalent to the star formation rate contributed by early and late type HS galaxies. For example, earlier type galaxies may host disc-mode star formation. To emphasize this, we also use cuts in the stellar disc fraction to calculate the star formation contributed by disc-dominated ($f^\star_{\rm D} \geq 0.5$) and spheroid-dominated ($f^\star_{\rm D} < 0.5$) HS galaxies.

Fig.~\ref{fig:sfrfrac} shows the fractional contribution of different components to the total star formation rate in bins of galaxy stellar mass and redshift. In both the top and bottom rows asymmetric galaxies are separated from HS systems using a stellar $A_{\rm 3D} > 0.2$ cut. The HS contribution is then split into disc and spheroidal gas structures (top row) or `earlier' and `later' type galaxies, via a stellar $f^\star_{\rm D}$ cut (bottom row). These panels may be compared to the stellar mass contribution of morphological components shown in Fig.~\ref{fig:frac}. 

Concentrating on the top row of Fig.~\ref{fig:sfrfrac}, we see that the SFR contributions from gas discs and spheroids differ markedly from the disc and spheroid stellar mass contributions of Fig.~\ref{fig:frac}. The most significant difference can be seen in the middle (spheroid) panel, where star formation in gaseous spheroids is negligible across redshifts, in all but the most massive galaxies. The low contribution to star formation by these structures is unsurprising, but a reassuring feature of the  morphological measurements.

As a result of the low star formation in spheroids, the panels for gas discs and asymmetric galaxies appear almost complementary. Generally asymmetric types dominate the star formation at high redshift. This then falls off towards lower redshift, but more gradually in lower stellar mass bins. By $z \sim 0$, discs dominate star formation over the entire mass range, and they dominate the $10 \lesssim \log_{10}(M^\star/{\rm M_\odot}) \lesssim 10.5$ range for $z \lesssim 2$. Again, high-mass asymmetric systems also contribute some significant star formation at $z > 0.5$, which could be attributed to merger remnants or star formation in massive galaxies with disturbed morphologies.

The fractional SFR contributions of `later' and `earlier' type HS galaxies in the bottom row of Fig.~\ref{fig:sfrfrac} are more balanced, revealing distributions closer to the respective disc and spheroid stellar mass panels of Fig.~\ref{fig:frac}. Taken together, this shows that while very little star formation occurs in gas spheroids, significant star formation takes place in gas discs belonging to in spheroid-dominated galaxies. 

\begin{figure}
\includegraphics[width=0.48\textwidth]{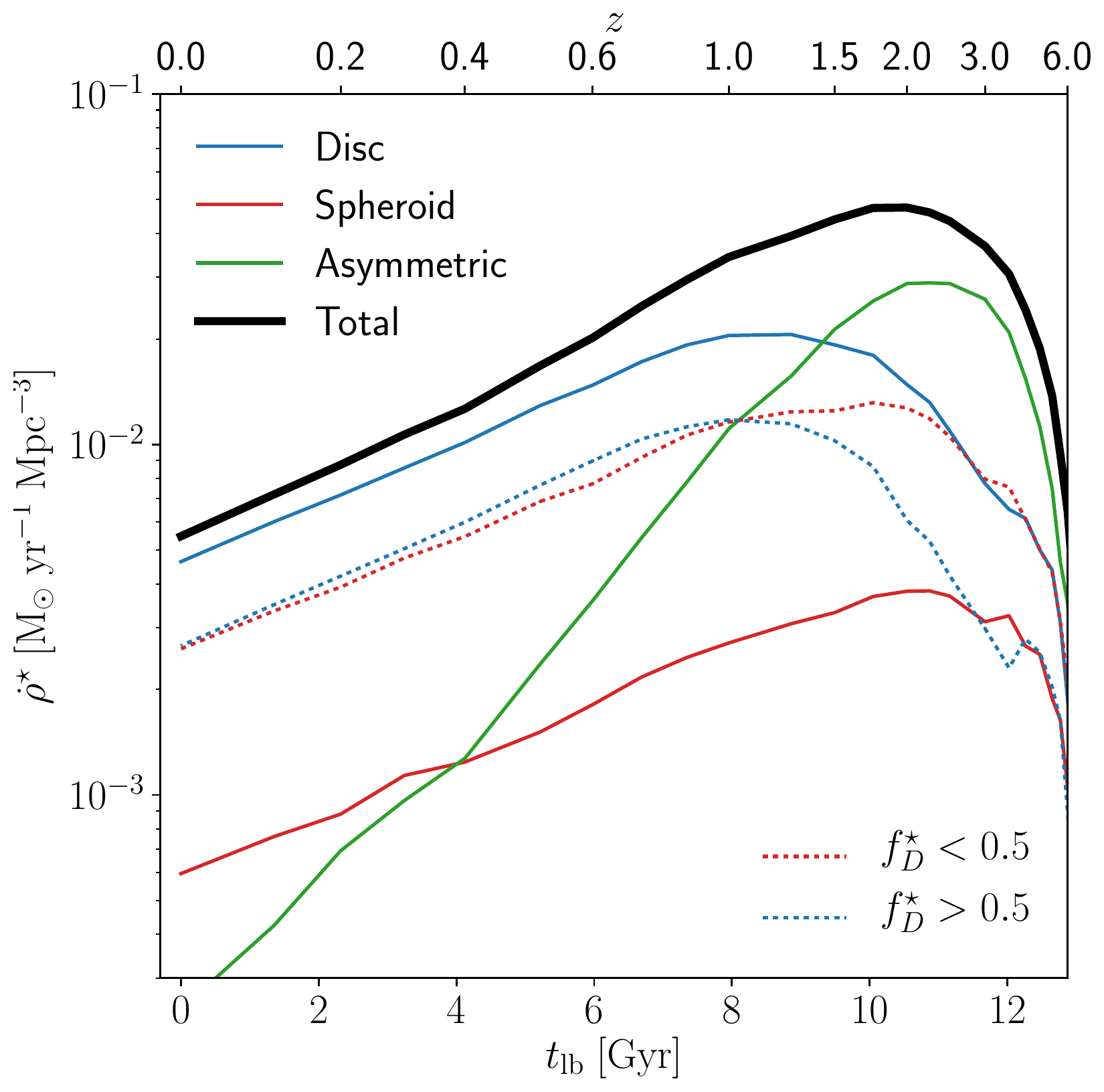}
    \caption{SFR density evolution for galaxies of $M^\star > 10^{9}{\rm M_\odot}$, split into the disc, spheroid and asymmetric component contributions. Asymmetric galaxies are defined to be above a stellar asymmetry threshold of $A_{\rm 3D} > 0.2$ (solid green). For the remaining HS galaxies, the spheroid contribution (solid red) is calculated as twice the SFR in counter-rotating gas, with the residual star formation assigned to the disc (solid blue). We also show the star formation rate contributions of HS galaxies with stellar masses dominated by spheroid and disc components (red- and blue-dotted lines respectively). Asymmetric galaxies dominate the cosmic star formation at high redshift, but are overtaken by gas discs in HS galaxies at $z\lesssim1$ to dominate the overall formation of stars. The comparable $f^\star_{\rm D} \geq 0.5$ and $f^\star_{\rm D} < 0.5$ lines suggest that gaseous disc structures found in galaxies with stellar masses dominated both by disc and spheroid components both contribute significantly to the overall SFR density.}
    \label{fig:cumsf}
\end{figure}

The evolving contribution of morphological components to the overall star formation rate density (SFRD) is shown in Fig.~\ref{fig:cumsf}. This can be compared to the cosmic stellar density contributions of Fig.~\ref{fig:cum}. Again, we split the contribution of HS galaxies into gaseous discs and spheroids, and into earlier and later type galaxies. A striking feature is that while gas discs contribute 85\% of the star formation at $z\approx 0$, discs harbor < 40\% of the stellar mass. Conversely, spheroids are the dominant component at $z\approx 0$ containing 60\% of the stellar mass, while spheroidal gas components contribute only 10\% of the cosmic SFRD. The dominance of discs is obscured when the HS star formation is split between earlier type ($f^\star_{\rm D} < 0.5$) and later type ($f^\star_{\rm D} \geq 0.5$) galaxies, with both contributing similarly at $z\approx 0$ (dotted lines).

Focussing now on the evolution of the SFRD for each component in Fig~\ref{fig:cumsf}, it is notable that the fractional contribution of gas discs increases monotonically from high redshift, despite their stellar mass contribution peaking at $z \approx 0.4$. In contrast, the near constant $\sim 10\%$ contribution to cosmic star formation by gas spheroids for $z \lesssim 3$ cannot account for the monotonic growth in the stellar mass contribution of spheroids through cosmic time, indicating that stellar discs must be transformed into spheroids. The evolving fractions of star formation and stellar mass in asymmetric systems are comparable, decreasing from 60\% at $z \approx 6$ to 6\% by $z \approx 0$. We find that it is asymmetric galaxies that contribute the most to $\dot{\rho}^\star$ at the peak of star formation, dominating at $z \gtrapprox 1.5$, but that stars formed in gas discs dominate the $z=0$ cosmic stellar mass density budget. 

\subsection{Growth and decay of Structures}
\label{sec:trans}

The difference between the stellar mass and star formation rate density contributions of disc, spheroid and asymmetric morphological structures are striking. They result from processes that transform the morphology of stellar systems after the stars are born, such as mergers and secular evolution. We refer to these collectively as \textit{transformational} processes. Here, we attempt to quantify the role of transformational processes for our sample of galaxies with $M^\star > 10^9 {\rm M_\odot}$.

Given that we have already computed the SFRs and stellar masses of disc, spheroidal and asymmetric structures, a simple approach would be to compare the integrated SFR associated with a given structural type to the overall mass growth rate of that type. The difference between these rates would be equivalent to the net rate at which transformational processes transfer mass into or out of that component, were it not for some subtleties. First, stellar mass loss through winds and supernovae means that, even without transformation, the overall mass growth of a system is less than its integrated SFR. Second, because we impose a minimum mass resolution below which galaxies are not tracked, galaxies growing above this threshold contribute to the mass growth of morphological components without contributing prior star formation.

To account for stellar mass loss, we compute the total stellar mass of a each component by summing the \textit{zero-age} main sequence masses of its stars, in addition to its present day stellar mass. The difference between the integrated star formation in, for example, asymmetric galaxies and the zero-age stellar mass of asymmetric galaxies today is then unaffected by stellar mass loss. To account for the lower mass limit below which we do not track galaxies, we compute the zero-age main sequence mass of structures in galaxies that have grown above the mass threshold between times $t_1$ and $t_2$ of successive snapshots.

Combining these, the change in (zero-age) mass in morphological component $X$ due to morphological transformations, $\Delta M^\star_{{\rm tran},X}$, is the difference in mass between $t_1$ and $t_2$ in this particular component, minus the change due to star formation, and minus the change due to galaxies entering the mass selection. The corresponding average transformation rate is 
\begin{equation}
\frac{\Delta M^\star_{{\rm tran},X}}{\Delta t} = \frac{\Delta M ^\star_X}{\Delta t} - \frac{{\rm SFR}_X(t_1) + {\rm SFR}_X(t_2)}{2}- \frac{\Delta M^\star_{{\rm enter},X}}{\Delta t} \; ,
\end{equation}
where, for morphological structure $X$, $M^\star_X$ is the total zero-age stellar mass at a given snapshot, ${\rm SFR}_X$ is the associated star formation rate, $M^\star_{{\rm enter},X}$ is the zero-age stellar mass associated with galaxies that have grown above the mass threshold, and  $M^\star_{\rm tran}$ is the zero-age stellar mass difference built up due to transformational processes. $\Delta$ indicates the difference in each quantity from $t_1$ to $t_2$, with $\Delta t = t_2 - t_1$.  We note that \textit{zero-age} mass growth rates are not directly observable. 

Using these rates, we can establish how transformational processes contribute to the overall mass growth or loss for each morphological component over cosmic time.  We show these as specific rates by dividing by $M^\star_X$, and refer specifically to the fractional growth rate of a component due to transformation as $\alpha_X$. 

We emphasise that the use of zero-age stellar masses in the calculation of growth rates allows us to separate the influence of stellar mass loss without tracking individual particles. The tacit assumption is that, on average, the stellar populations transferred between components are at the same phase of their mass loss. While typical ages clearly differ between components, this assumption is well motivated because most of the stellar mass loss is due to evolution of massive stars, which have very short lifetimes \citep[$< 100$~Myr after formation e.g.][]{Wiersma09b}. such that only the youngest stellar populations have retained a significantly higher fraction of their zero-age mass. Indeed, we find that the overall mass loss fractions\footnote{Found by dividing the present-day mass in $X$ components by their zero-age mass.} for each component remain within $\approx 5$~\% of each other at all times. However, we note that this differs from the standard presentation of specific star formation rates (sSFR), where the rate of mass formation of an object is normalised by the current mass. As a consequence, the growth rates due to star formation that we derive for each component are $\approx 60$~\% of the sSFR at $z=0$ and $\approx 65\%$ at $z=2$. 

\begin{figure*}
\includegraphics[width=0.98\textwidth]{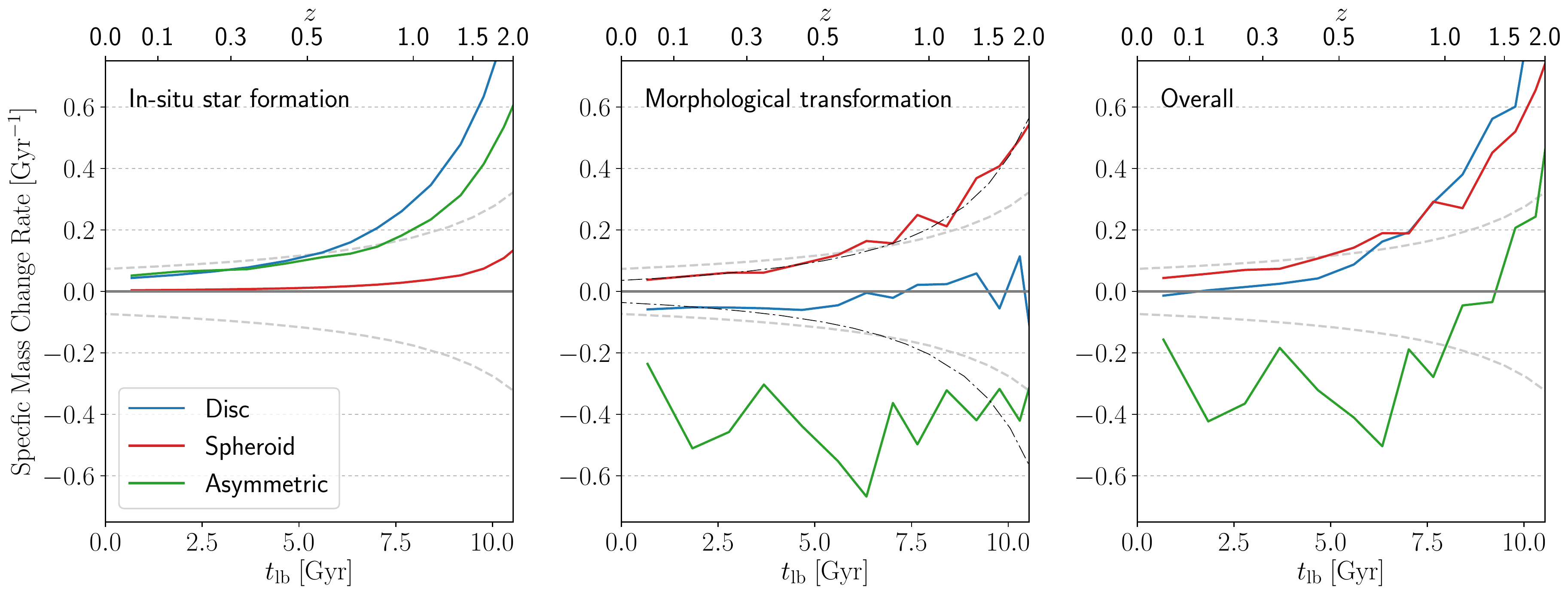}
    \caption{Specific rates of change in stellar mass for different morphological components (see text for details) in galaxies of $\log_{10}(M^\star/{\rm M_\odot} > 9)$. We plot the specific mass change rate contributed by \textit{in-situ} star formation (left) and morphological transformation of the stellar structures (middle). The net rate is plotted for comparison (right). Dashed lines indicated $\pm 1/t_{\rm age}$, so that growth or decay timescales can be compared to the Hubble time. Dot-dashed lines represent a model redshift dependence for the galaxy merger rate, $\propto (1+z)^{2.5}$ (see text), normalised to best fit the bulge growth rate at $z<2$ and mirrored in the x-axis. In calculating the different rates, we have separated the specific mass loss due to stellar evolution, and the spurious effect of galaxies growing above the minimum mass threshold. Both contribute marginally to the net rate at low redshift (see text). We see that spheroid growth is driven by subsuming stars from other components, while the early build-up and eventual decay of discs is driven by \textit{in-situ} star formation giving way to loss of disc stars to other components.}
    \label{fig:specrate}
\end{figure*}

In Fig.~\ref{fig:specrate} we show the specific growth rates associated with star formation, morphological transformation and all processes. We first inspect how the star formation contribution to the growth rate evolves for each component. We see that all of these rise monotonically with lookback time. For $0.3\lesssim z \lesssim 2$ the star formation growth rate associated with discs is highest, with asymmetric morphologies second and spheroids a distant third. At $z\lesssim 0.3$ the star formation growth rate in asymmetric galaxies is slightly higher than that of discs. By comparing to the Hubble timescale (dashed) lines, we see that at $z<0.5$ the growth timescale associated with star formation becomes longer than the Hubble time. For discs and asymmetric systems, this timescale remains comparable to the Hubble time, while  for spheroids the timescale is more than three times the Hubble time at $z < 1.5$. 

We now turn to the rates associated with transformational process for each component at $z \lesssim 2$, shown in the middle panel of Fig.~\ref{fig:specrate}. We see that spheroids have the highest transformation growth rate ($\alpha$) at all times for $z < 2$, and this dominates their growth due to star formation. The plot shows how spheroids are built primarily by subsuming stellar material from other structures. However, the growth timescale for spheroids is close to the Hubble time for $z \lesssim 0.5$. Conversely, asymmetric galaxies are  depleted by transformational processes. While the transformation rate is stochastic, it shows no clear trend with redshift at $z<2$ and varies around a decay rate $\sim -0.4$~Gyr$^{-1}$. This leads to an overall decay in the mass in asymmetric systems for $z < 1.5$ (right panel). Discs are also net destroyed by transformational processes at late times, at a near constant rate of $-0.07$~Gyr$^{-1}$ at $z<0.5$. The positive growth at $z > 0.5$ is dominated by star formation. As specific star formation in discs drop, there is a very slight decay in the mass of disc structures by late times ($z<0.1$). 

The constancy of the decay rate in discs at $z<0.5$ gives some clues to the mechanisms driving transformation. Mergers are often invoked as mechanisms of disc destruction. The behaviour of the galaxy merger rates in hydrodynamical simulations depend only weakly on mass ratio and is similar to that of the halo merger rate obtained from N-body simulations \citep{Genel10,Fakhouri10, RodriguezGomez15}, with a strong redshift dependence of around $(1+z)^{2.5}$. If mergers were to dominate the destruction of discs, we might reasonably expect the disc destruction rate to be proportional to the merger probability, which itself deceases by a factor of 5.7 from $z=1$ to $z=0$. A complication of this picture is demonstrated by \citet{Lagos18}: the angular momentum of merger remnants in EAGLE depends somewhat on the gas fraction of mergers which itself changes with redshift (see their Fig.~1). However, the factor $\approx 2$ increase in the fraction of gas-free mergers between $z=1$ and $z=0$ cannot balance the factor $\approx 5$ decrease in mergers overall. The lack of evolution in the disc destruction rate is also consistent with the finding by \citet{Clauwens17} that mergers  contribute to both the destruction and growth of discs, with little net effect.

In the case that disc destruction is dominated by secular processes (e.g. instabilities), a proportionality can be assumed between the mass transfer rate from disc to bulge and the mass in discs. This is then compatible with the constant specific decay rates we measure for discs. While secular processes may dominate the mass transfer between discs and spheroids in EAGLE, mergers may still supplement or even dominate the growth of spheroids, as found by \citet{Clauwens17}.

Generally, the morphological transformation rates of each structure we identify show distinct evolution, albeit with significant stochastic variability. Some of this variation can be attributed to the definition of asymmetric galaxies, as the mass in a galaxy crossing the somewhat \textit{ad-hoc} threshold of $A_{\rm 3D} = 0.2$ shifts all its mass between the disc/spheroid and asymmetric contributions. The effect of the exact shape criterion for asymmetric systems is explored further in appendix \ref{ap:pecs}.

It is important to note that measuring the growth and decay of discs depends on how we define disc structures. For instance, the \textit{prograde excess} that we use to define disc structures merely requires that disc material be co-rotating, and makes no additional requirement on the dynamical temperature of this material. We further explore the disc definition and the evolution of circularities in the following section.

\section{Evolution of circularity distributions}
\label{sec:micro}

So far our morphological analysis has focused on integrated galaxy properties. Disc fractions, moments of inertia and asymmetry values are computed from stars and star-forming gas on a galaxy-by-galaxy basis. However, considering the integrated properties alone could miss subtleties in the dynamical evolution of galaxies. We can take advantage of the mass resolution of EAGLE to explore further the internal dynamics of the simulated galaxies using the individual SPH particles.

As mentioned in Section \ref{sec:kins}, we compute the circularity values, $\epsilon$, for individual star and gas particles following \citet{Abadi03}:
\begin{equation}
\epsilon = j_{\rm z}/j_{\rm circ}(E),
\end{equation}
where $j_{\rm z}$ is the component of specific angular momentum along the primary rotation axis, and $j_{\rm circ}(E)$ is the total specific angular momentum of a circular orbit of equivalent energy. This energy is computed by estimating the gravitational potential from the halo mass profile out to the virial radius, $R_{200}$, assuming sphericity. The stellar $\epsilon$ distribution for a classical disc peaks at $\epsilon \approx 1$, while a classical spheroid shows a symmetrical peak at $\epsilon \approx 0$.

\subsection{Stellar circularities at birth}
\label{sec:sfcirc}

In the previous sections, the spheroid mass was estimated by doubling the counter-rotating mass, with the remaining \textit{prograde excess} representing the disc. While the individual $\epsilon$ values are not needed to compute this, they can be used to characterise features such as slowly rotating spheroidal components or psuedobulges \citep[e.g.][]{Scannapieco09}, dynamically hot discs, or even counter-rotating structures. We can also use circularities to include dynamical temperature criteria in the definition of discs, and see how this affects the evolution of disc fractions. 

\begin{figure}
\includegraphics[width=0.48\textwidth]{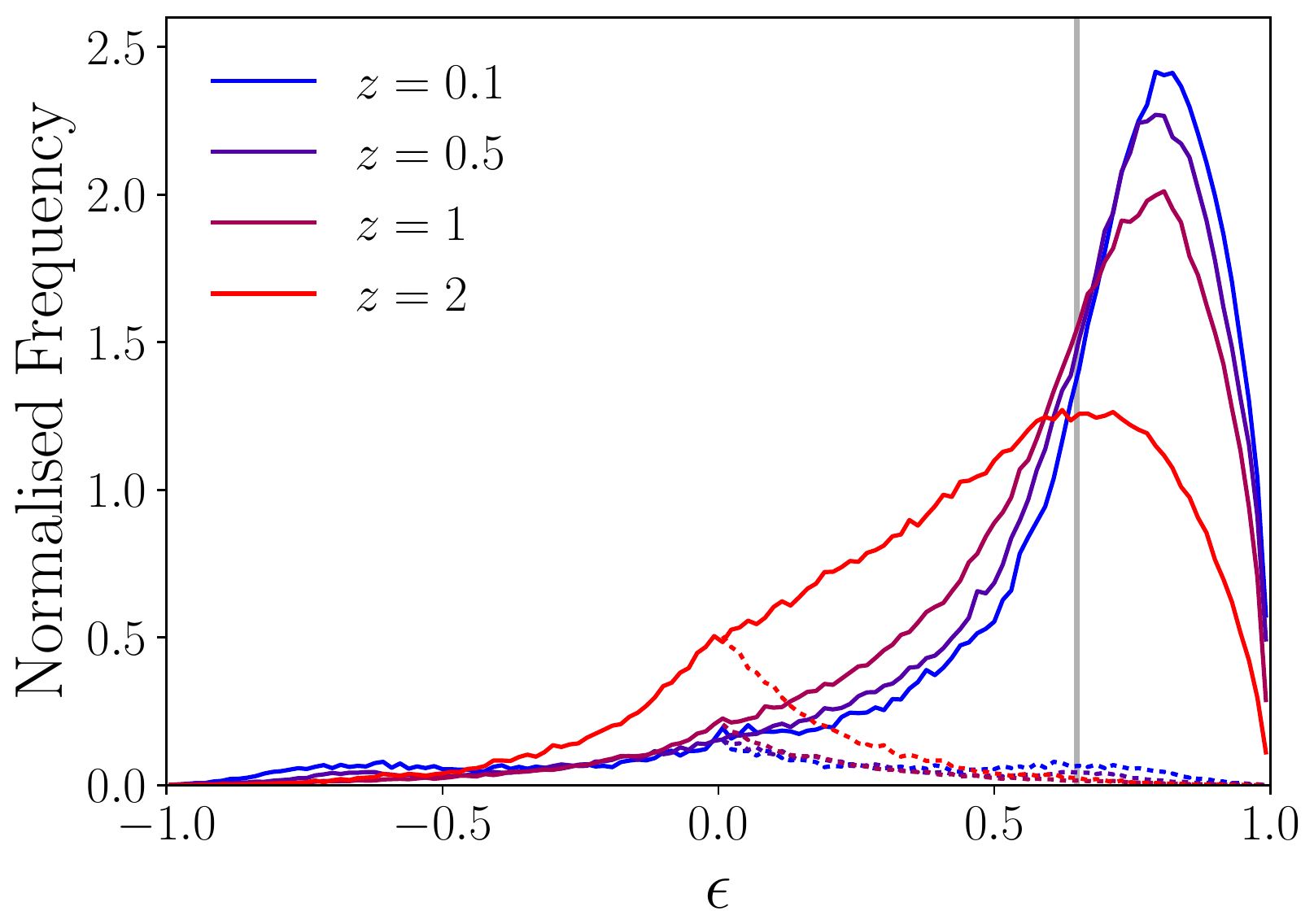}
    \caption{SFR-weighted circularity distributions of gas particles taken from HS galaxies at redshifts $z=0$, 0.5, 1 and 2. Particles are selected randomly from well resolved galaxies in the Ref-100 volume (with $\log_{10}(M^\star/{\rm M_\odot) > 10}$), with a down-sampling frequency of $10^{-3}$. Dotted lines reflect the negative $\epsilon$ distribution about $\epsilon=0$, illustrating the case of no prograde excess, while the grey vertical line shows the $\epsilon > 0.65$ cut, used as an alternative definition of disc material. We see clear evolution to dynamically cooler star-forming gas with decreasing redshift, changing from a broad peak at $\epsilon \approx 0.65$ at $z=2$ to a narrower peak at $\epsilon\approx 0.8$ by $z=0$.}
\label{fig:sfeps}
\end{figure}

Fig.~\ref{fig:sfeps} shows the SFR-weighted $\epsilon$ distributions of star-forming gas particles at redshifts 0, 0.5, 1 and 2. These are constructed from gas particles within 30~pkpc of the center of HS (i.e. $A_{\rm 3D} > 0.2$) galaxies with $\log_{10}{M^\star/{\rm M_\odot}} > 10$, randomly downsampled by a factor of 1000 to reduce computational expense. Clear evolution of the distribution can be seen with redshift. The height of the narrow peak (FWHM$\approx0.4$) at $\epsilon\approx0.8$ in the $z=0$ distribution falls with increasing redshift, while star formation in gas particles with  $\epsilon \lessapprox 0.8$ increases monotonically. At $z=2$ the peak has fallen to $\epsilon\approx 0.65$ with a FWHM of $\approx 1$. While co-rotating gas dominates star formation across all epochs, the star formation generally takes place in dynamically hotter systems at higher redshifts. Such evolution of star-forming gas kinematics has previously been found in cosmological zooms of individual disc systems \citep[e.g.][]{Brook12, Ma17, Navarro17}, and is consistent with the observational finding that gas kinematics show less rotational support at high redshift \citep[see e.g.][]{Tacconi13, Swinbank15, Turner17}.

\begin{table}
  \caption{The disc fraction for stellar mass and SFR in HS galaxies of $M^\star > 10^{10} {\rm M_\odot}$, using two definitions of the disc component. At four redshifts (first column) the disc fraction is calculated either as the \textit{prograde excess}, which defines $f_D$ for $M^\star$ and SFR in this work, or by a minimum circularity cut of $\epsilon > 0.65$ \citep{Tissera12}. We see that the $\epsilon > 0.65$ fractions capture less SFR and $M^\star$ than the prograde excess. The prograde excess fractions show little evolution, while the  $\epsilon > 0.65$ fractions show larger changes.} 
\label{tab:discs}
\centering 
\begin{tabular}{l | c c | c c} 
\hline\hline 
Redshift & \multicolumn{2}{|c}{Prograde excess fraction} & \multicolumn{2}{|c}{$\epsilon > 0.65$ fraction}\\ [0.5ex]
$z$ & SFR & $M^\star$ & SFR & $M^\star$ \\
\hline 
0.1 & 0.88 & 0.42 & 0.66& 0.20\\ 
0.5 & 0.90  & 0.47 & 0.62& 0.21\\
1 & 0.91 & 0.49 & 0.55& 0.32\\
2 & 0.79 & 0.41 & 0.33& 0.14\\ [0.5ex] 
\hline 
\end{tabular}
\end{table}

Fig.~\ref{fig:sfeps} highlights the importance of disc definition in the recovered disc fractions. To illustrate this, the disc fraction is calculated for each redshift using the \citet{Abadi03} approach adopted in this work, and the more direct cut of $\epsilon>0.65$ used by \citet{Tissera12}. These values are tabulated in Table~\ref{tab:discs}. We see that the prograde excess dominates star formation, with star formation associated with the spheroid remaining at a relatively constant low fraction for $z\lessapprox 1$, increasing from 9\% to 21\% between $z=1$ and 2. In contrast, pronounced evolution in the profile of the positive $\epsilon$ peak causes disc fractions recovered with the $\epsilon > 0.65$ cut to change more dramatically. In this case, star formation changes from being disc-dominated ($f_{D,\;{\rm SF}} = 0.66$) at $z=0$ to spheroid dominated ($f_{D,\;{\rm SF}} = 0.33$) at $z=2$. This disparity shows (for $z<2$) that while gas on prograde orbits overwhelmingly dominates star formation, this star-forming component becomes increasingly dynamically cold with cosmic time; high-redshift gas `discs' show messier kinematics.

\begin{figure}
\includegraphics[width=0.48\textwidth]{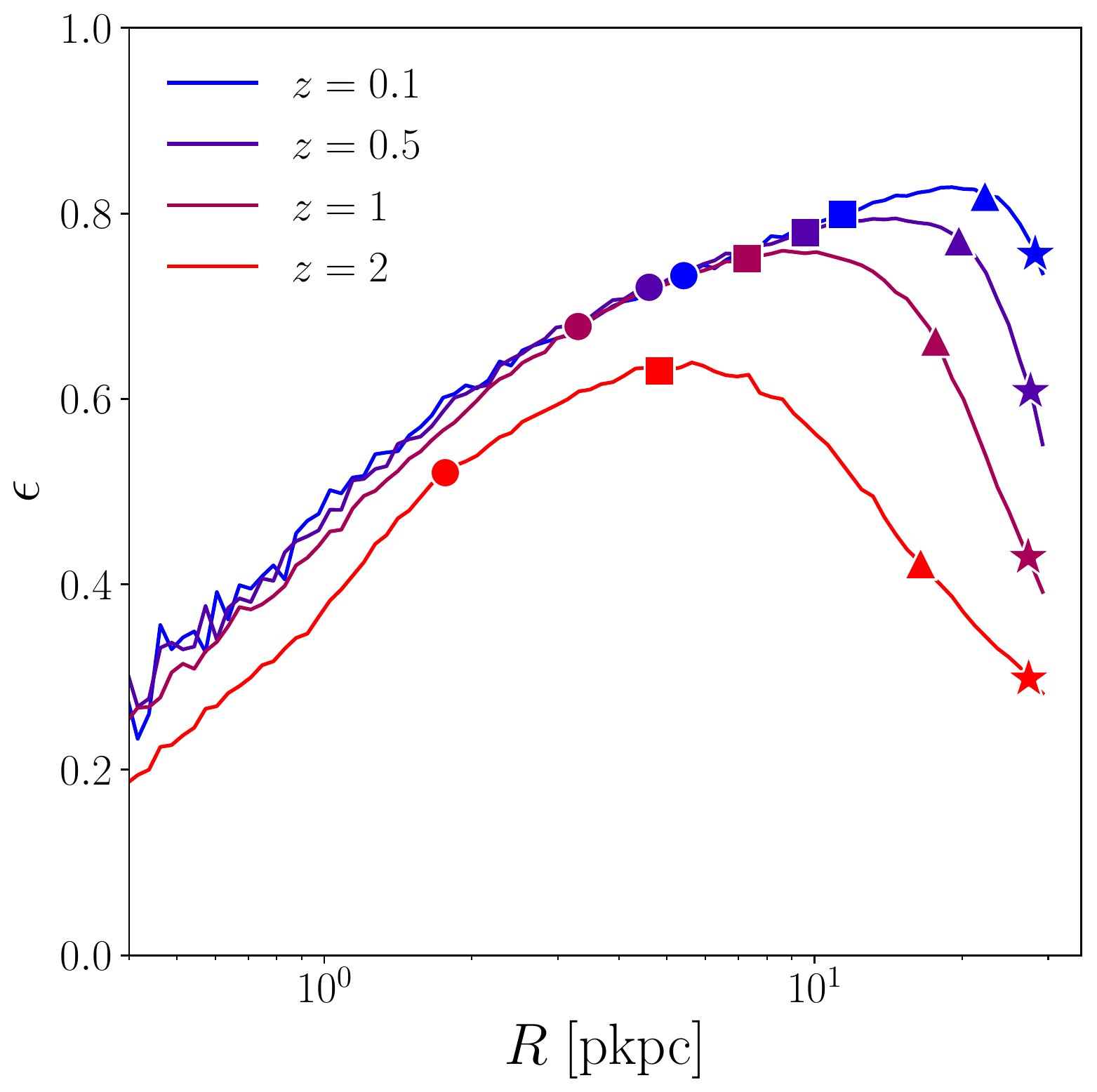}
\caption{Median circularity of star-forming gas in $\log_{10}(M^\star/{\rm M_\odot) > 10}$ galaxies as a function of radius. These are split by redshift, using the same particle selection and redshift values as in Fig.~\ref{fig:sfeps}. Markers indicate the radii enclosing 50\%, 75\%, 95\% and 99.5\% of the total SFR (circles, squares, triangles and stars respectively) coloured to correspond to each $\epsilon$ profile. We see that while star formation is more central at higher redshift, the circularites are generally lower at all radii with increasing redshift. We also note an evolving radius at which the median circularities peak, such that typical circularities at $R \gtrsim 10$~kpc differ dramatically between different redshifts.}
    \label{fig:radial}
\end{figure}

It is difficult to say from this alone what is driving the evolving dynamics of the star-forming gas. While it may reflect true evolution in the kinematics of galaxies with redshift, it is feasible that this change is instead driven indirectly by the evolution of other galaxy properties. In appendix~\ref{ap:enviro} (Fig.~\ref{fig:enviro}) we investigate the influence of the evolving  mass distributions and typical environments of galaxies with redshift. We find that trends with mass and environment are detectable, but that this alone cannot explain the evolution as comparable redshift trends persist for fixed bins of stellar and halo mass.

Beyond stellar mass and environment, A third property that shows strong redshift evolution and could influence circularities is galaxy size. For example, if there exists a radial trend in gas circularity with galactocentric radius, the more compact nature of higher-redshift galaxies could drive the evolution in Fig.~\ref{fig:sfeps}, even without evolution in the radial trend itself. Such a trend could be physical in origin but could also be influenced by numerical artefacts, particularly when the radius is comparable to the gravitational smoothing. Fig.\ref{fig:radial} shows the median circularity values as a function of radius for star-forming gas at different redshifts, using the same particle selection and redshifts as in Fig.~\ref{fig:sfeps}. Markers show the radii enclosing 50\%, 75\%, 95\% and 99.5\% of the total SFR on each line from left to right. We see that in the inner parts there is indeed a strong increasing trend of circularity with radius, and that the majority of star-forming gas is generally more concentrated at higher-redshift, particularly for the 50th and 75th percentile radii. While this partly contributes to the lower circularity of star-forming gas at higher redshift in Fig.~\ref{fig:sfeps}, the higher-redshift circularities are generally lower at all radii. This is particularly pronounced at $z=2$. The messier kinematics of high redshift galaxies likely influences the increasing prominence of asymmetric galaxies towards higher redshift, explored further below.

An intriguing feature of the median circularity profiles is the behaviour at larger radii. A turnover radius is observed for each redshift at which the relation changes from increasing to decreasing circularity with radius. The radius of this peak evolves by a factor of 4 from $R \approx 5$~pkpc at $z=2$ to $R \approx 20$~pkpc by $z=0.1$, leading to dramatically different dynamics for the most extended 25\% of star formation at different redshifts. It is unclear what drives the turnover, but the influence of disc flaring or of a prograde rotating stellar halo component in the outer parts could cause $\epsilon$ to drop at high radii.

Overall, we find an increase in the dynamical temperature of star forming gas with redshift which cannot be explained by stellar mass, environmental and size trends alone. If discs are required to be sufficiently dynamically cold, rather than merely a prograde excess, this leads to stronger evolution in the fraction of star formation in discs with redshift. This demonstrates how a characterisation of the morphological mix and evolution of galaxies is influenced by the disc definition. In what follows we consider how stellar circularity distributions differ from that of their natal gas, and the relevant transformational processes.

\subsection{Stellar circularities with age}
\label{sec:starcirc}

\begin{figure}
\includegraphics[width=0.48\textwidth]{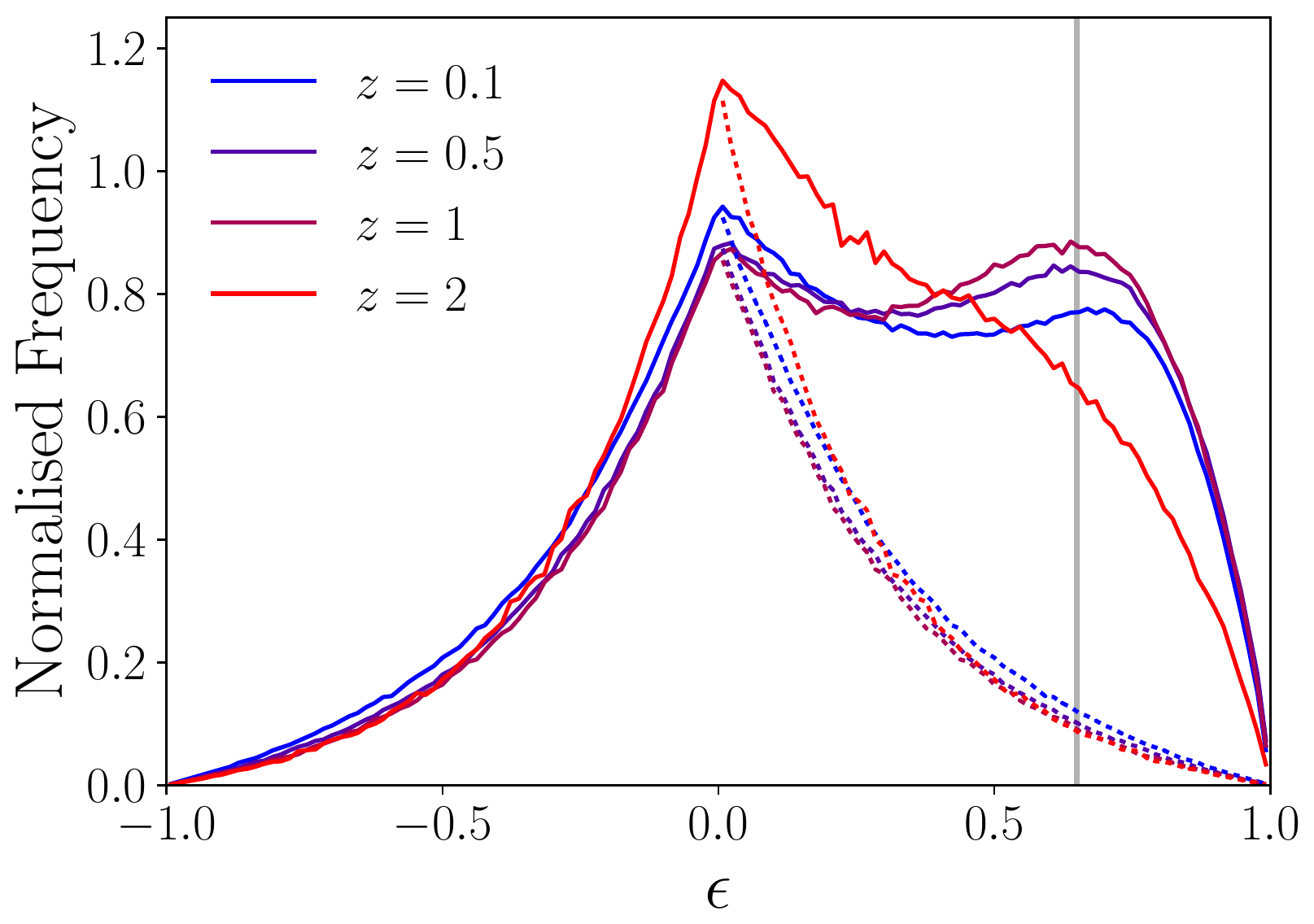}
    \caption{As Fig.~\ref{fig:sfeps}, but now showing the stellar mass weighted circularity distributions. A bimodality is seen in the stars that is not apparent in the star-forming gas. The prograde peak shifts to lower $\epsilon$ values with increasing redshift.}
\label{fig:stareps}
\end{figure}

\begin{figure}
\includegraphics[width=0.48\textwidth]{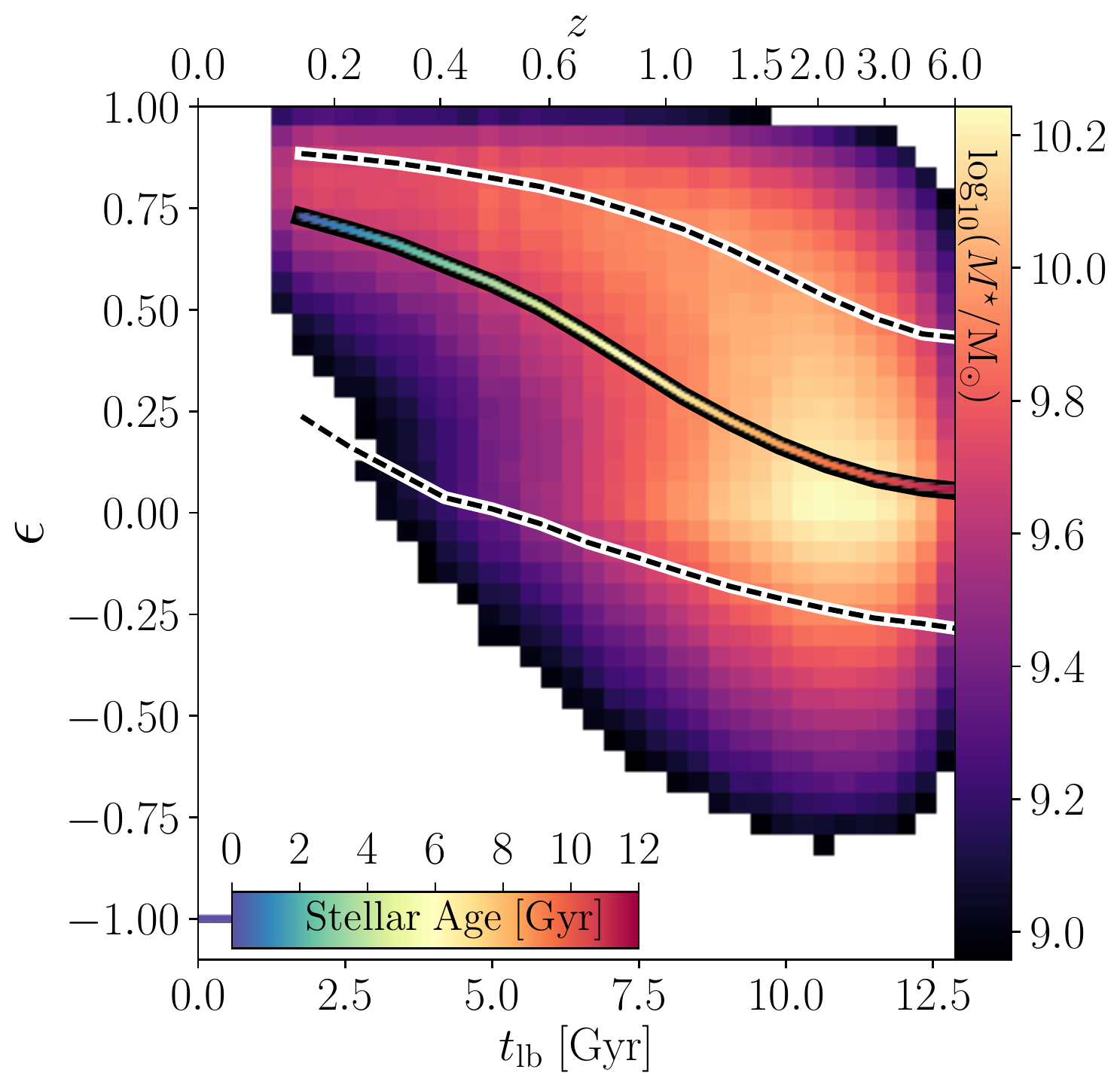}
    \caption{The mass-weighted circularity distribution of stars as a function of stellar age (lookback time to formation) at $z=0.1$ for $\log_{10}(M^\star/{\rm M_\odot) > 10}$ galaxies. The number density of stars in each $\epsilon$-$t_{\rm lb}$ bin is indicated by the underlying colour map. The median $\epsilon$ value as a function of $t_{\rm lb}$ is shown by the solid line, with dashed lines indicating the 16$^{\rm th}$-84$^{\rm th}$ percentile range. The stars show a trend towards more spheroid-like kinematics with age, but the distribution of $\epsilon$ is bimodal for $4 \lessapprox t_{\rm lb}/{\rm Gyr} \lessapprox 8$.}
    \label{fig:eps}
\end{figure}

Fig.~\ref{fig:stareps} is analogous to Fig.~\ref{fig:sfeps}, but now weighted by stellar mass at each redshift, while Fig.~\ref{fig:eps} shows the bivariate distribution of circularity vs age for stellar populations found in \textbf{$z=0.1$} galaxies. Again, we use galaxies with $\log_{10}(M^\star/{\rm M_\odot}) > 10$. 

Comparing the stellar mass weighted $\epsilon$ distribution of Fig.~\ref{fig:stareps} to the SFR-weighted distribution of  Fig.~\ref{fig:sfeps} is informative. We see less prominent discs and less variation in the global disc/spheroid ratios for the stellar component, as quantified in Table~\ref{tab:discs}. The $\epsilon > 0$ portion of the stellar distributions account for less material and peak at systematically lower $\epsilon$ values than the corresponding gas. This is a consequence of the transformational processes discussed in section \ref{sec:trans} and the timescales over which they act. Inspecting circularity as a function of stellar age allows us to explore this further.

In Fig.~\ref{fig:eps} the median circularity for a given stellar age is plotted as the shaded line. For young stars ($\lessapprox 2$ Gyr old) the stellar circularities trace the $z=0.1$ star-forming gas distribution, shown in Fig.~\ref{fig:sfeps}, with a single peak at $\epsilon \approx 0.8$. However, the underlying circularity distribution for intermediate-age stars, born at $z\approx 0.5$ ($t_{\rm age} \approx 4.2$~Gyr), shows a secondary peak at $\epsilon=0$, which is absent from the progenitor gas distribution in Fig.~\ref{fig:sfeps}. This is indicative of  a distinct spheroid component, which is assembled after the stars were born. The spheroid component becomes more significant with age, with stars older than 9~Gyr showing little prograde excess. 

\begin{figure}
\includegraphics[width=0.48\textwidth]{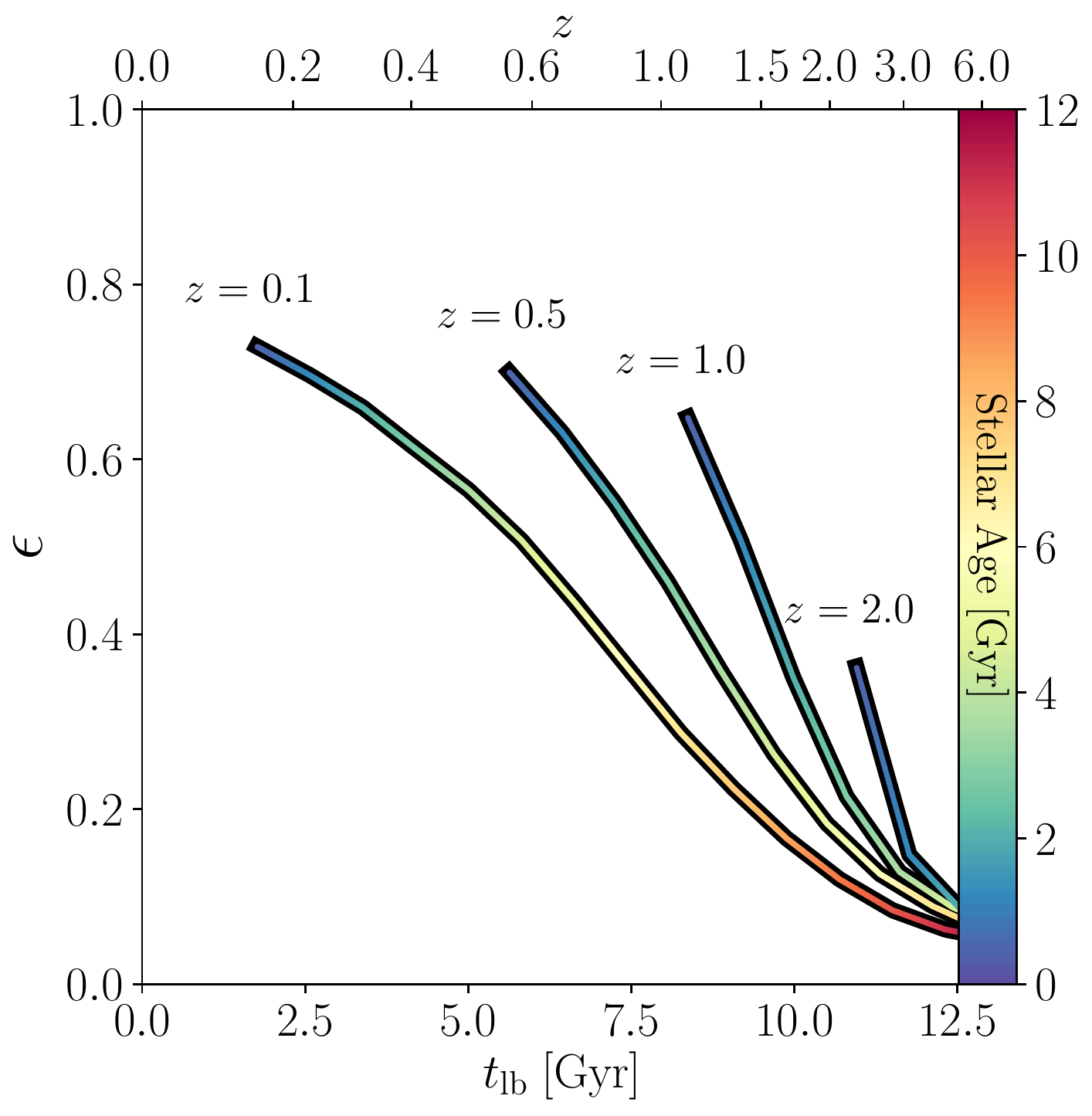}

\caption{Circularity of stars and measured at different redshifts for $\log_{10}(M^\star/{\rm M_\odot) > 10}$ galaxies. \textit{Lines} show the median circularity as a function of the lookback time to when the stars formed, coloured by the age of stars as indicated by the color bar. Different lines correspond to different redshifts from which the $\epsilon$-age relation of stars is measured; originating at 0.1, 0.5, 1 and 2 from left to right. In the absence of evolution, these lines would overlap. We see that the low-redshift circularity distribution of stars is influenced by both evolution in the typical circularity of the stars at formation (i.e. the leftmost ends of each line) and transformational evolution of stars after birth (difference between lines at a given $t_{\rm lb}$).}
    \label{fig:epsevol}
\end{figure}

Considering the individual particles provides further insight into how  both the evolution of gas dynamics in galaxies and processes acting on existing stars shape the morphologies of EAGLE galaxies. Fig.~\ref{fig:epsevol} illustrates both aspects, using the age-dependent median circularities of stars selected at each of $z=0$, 0.5, 1, 2. These are plotted as a function of lookback time from $z=0$, so that stellar populations of the same age and those born from the same gas can be compared at different snapshots. 

The leftmost point of each line reflects the median circularities of stars at birth. As seen in Fig~\ref{fig:sfeps}, more stars are born with lower circularities at higher redshift. However, this only marginally changes (by $\approx15\%$) the medians for the youngest stars and star-forming gas at $z\lessapprox1$, with a more pronounced change (by $\approx 50\%$) from $z\approx2$. This suggests that the globally higher spheroid fractions for old stars measured at $z=0.1$ ($\gtrapprox$10~Gyr, born at $z\gtrapprox2$) may be significantly influenced by the lower birth circularities, but the presence of a younger spheroid component is further evidence for transformational processes acting over the lifetime of stars. 

The effects of transformational processes on individual star particles can be seen directly by comparing the median stellar $\epsilon$ vs age relations for different snapshots at a given lookback time in Fig.~\ref{fig:epsevol}. For example, the stars born at $t_{\rm lb} \approx 5.5$~Gyr from $z=0.1$ have median circularities of $\epsilon \approx 0.75$ in their newborn state at $z=0.5$, but evolve to circularities of $\epsilon \approx 0.5$ by $z\approx0.1$. The rate at which transformational processes change the stellar morphologies was demonstrated to evolve with redshift in Fig~\ref{fig:specrate}. Comparing the stars of the same age at different snapshots, via the shading of each line, shows the evolving rate at which $\epsilon$ values are transformed with redshift. Comparing coeval stellar populations at consecutive redshifts, we see that at $z=0.1$ the median $\epsilon$ value for the stars born at $z\approx0.5$ falls by $\approx 15\%$ (from $\epsilon \approx$ 0.7 to 0.5) over a $\approx 4$~Gyr period. At $z=0.5$ the stars born at $z\approx1$ have changed more dramatically from their birth values, by $40\%$ (from $\epsilon \approx$ 0.65 to 0.4) over a shorter $\approx 3$~Gyr period.

We glean a number of important insights from investigating the circularity distributions of stellar and star-forming particles in EAGLE. While the circularities of star-forming gas  are generally lower at higher redshift, transformational processes are necessary to build a distinct spheroid component; the bimodal distribution of stellar circularities is absent for the star-forming gas. The spheroidal stellar component peaks at $\epsilon \approx 0$, suggesting that rotation of the spheroid \citep[e.g.][]{Scannapieco09} is of little importance globally and that taking twice the retrograde rotating material is a reasonable proxy for the spheroid. The evolving circularity of star-forming gas introduces ambiguity in the definition of a disc; while the star-forming gas is overwhelmingly prograde out to high redshift ($z \lessapprox 2$), the gas is dynamically hotter at higher redshift. The `\textit{discs'} identified at high redshift using the prograde excess are clearly of a different character to those at $z\approx0$. `Messier' dynamics likely also account for the higher proportion of asymmetric galaxies found at all masses at high redshifts ($z \gtrapprox 2$). 

\section{Summary \& Conclusions}
\label{sec:conc}

We have investigated the evolving morphological properties of EAGLE galaxies, using the three dimensional mass distributions and kinematics of baryonic material. In this work, morphologies are measured in the physical domain, as a prelude to future studies utilising mock observations. In particular, we consider morphological evolution by defining and following three types of morphological structures: disc, spheroid and asymmetric systems. The contribution of these structures to the cosmic stellar mass and star formation rate budget through cosmic time are then quantified, such that the evolving prominence of different structures can be assessed.

In order to define the three components, we first separate galaxies into those conforming to the `Hubble Sequence' from those with more irregular morphologies, by inspecting their shape and asymmetry. We find that Hubble types are not cleanly separated by their axial ratios, spanning a range of shapes from oblate late types to spherical and prolate early types. We instead define a measure of 3D asymmetry, and use a threshold value $A_{\rm 3D} > 0.2$ to separate Hubble types from disturbed morphologies (see Fig.~\ref{fig:a3ds}). The $A_{\rm 3D} > 0.2$ galaxies are deemed \textit{asymmetric} structures in our nomenclature.

Having designated Hubble-type galaxies, we quantify their spheroid-to-total ratios for stellar mass (or for star formation rate) using the particle orbits. This follows the approach of \citet{Abadi03}, where twice the counterrotating mass (SFR) is taken to be the total mass (SFR) contribution of the spheroid. The residual mass (SFR) is then taken to reside in the disc, attributable to the excess prograde material. We note that this definition of the disc is rather broad, and merely imposes that disc material must be corotating. More stringent cuts based on orbital circularities are explored in Section \ref{sec:micro}. 

From analysing the mass and redshift distributions of disc, spheroid and asymmetric structures within EAGLE galaxies of $M^\star > 10^{9}{\rm M_\odot}$, we obtain a number of key findings:  
\begin{itemize}
\item Asymmetric morphologies dominate at high redshift, particularly at lower stellar masses (Fig.~\ref{fig:frac}). At high redshift this is taken to reflect that asymmetric systems are just being assembled and relaxation processes have not had time to redistribute stellar orbits.
\item A galaxy stellar mass of $M^\star \sim 10^{10.5}{\rm M_\odot}$ (corresponding to a halo mass of $M_h \sim 10^{12}{\rm M_\odot}$) is the characteristic scale at which the mass contribution of discs peaks for $z \lessapprox 3$ (Fig.~\ref{fig:frac}).
\item Hubble sequence galaxies dominate the stellar mass at $z\lessapprox 1.5$, with spheroids the dominant morphological structure by mass (Fig.~\ref{fig:cum}). While the mass fraction of spheroids increases steadily towards the present day,  discs reach a peak mass contribution at $z \approx 0.5$.
\item The majority of stars were formed in gaseous discs, despite discs always containing a subdominant fraction of the cosmic stellar mass density (Figs.~\ref{fig:cum} and \ref{fig:cumsf}). The fraction of star formation in gaseous discs increases with cosmic time, overtaking star formation in asymmetric systems at $z < 1.2$.
\item Kinematically spheroidal gas contributes the lowest fraction of star formation at all redshifts (Fig.~\ref{fig:cumsf}), demonstrating the importance of transformational processes that lead to spheroids dominating the cosmic stellar mass.
\item While spheroids are built up by subsuming stellar mass from other structures, transformational processes lead to the removal of stars from discs at an approximately constant specific rate of $\sim 0.07$~Gyr$^{-1}$ at $z\lessapprox 0.5$ (Fig.~\ref{fig:specrate}). 
\item The constancy of the disc destruction rate (Fig.~\ref{fig:specrate}) may imply that, on a cosmic scale, disc destruction is dominated by secular processes, and largely independent of the cosmic merger rate which increases strongly towards higher redshifts. For individual galaxies, however, mergers play an important role in wholesale morphological transformation. Following individual galaxies, we find that those that change from significantly disc-dominated to spheroid-dominated (Fig.~\ref{fig:bars}) show a prevalence of mergers with stellar mass ratios of $\gtrapprox 0.1$ relative to the overall population. The spheroid growth rate also exhibits a strikingly similar redshift dependence to the galaxy merger rate.    
\end{itemize}

It is important to note that our quantitative findings are contingent on the way in which different morphological structures are defined. In particular, the choice of the Hubble sequence asymmetry criterion ($A_{\rm 3D} < 0.2$) is somewhat \textit{ad-hoc}, and defining the prograde excess to be the disc merely requires that the orbit be corotating with the net angular momentum vector. While changing these definitions will change the quantitative results, the general qualitative picture of morphological evolution that arises from those we use is robust.

For insight into the limitations of these definitions, we investigated the orbital \textit{circularity} of baryons in EAGLE galaxies in Section~\ref{sec:micro}, and found the following:
\begin{itemize}
\item The star-forming gas discs we define are dynamically colder at low redshift than those at high redshift (Fig.~\ref{fig:sfeps}).
\item Transformational processes are needed to build up a bimodal circularity distribution of stars. While lower birth circularities at high redshift contribute to lower circularities of older stars, they do not account for the peak at $\epsilon \approx 0$ (Figs.~\ref{fig:stareps} and \ref{fig:epsevol}).
\item Defining discs as the prograde excess is less sensitive to evolution in rotational dynamics than applying a minimum circularity threshold to define a disc, which requires disc material to be sufficiently dynamicaly cold as well as corotating. We see a more dramatic evolution in the disc fraction at $z<2$ using a circularity cut, with much lower disc fractions at high redshift (Table~\ref{tab:discs}).
\end{itemize}

We find a coherent picture of morphological evolution in EAGLE that builds upon the three phase model of \citet{Clauwens17}. Using the spheroid-to-total ratios\footnote{Defined as twice the couterrotating stellar mass.} to measure morphology, they find that morphological change is primarily a function of galaxy stellar mass and is remarkably independent of redshift for $z < 3$. This is largely supported by our findings for galaxies with $M^\star \gtrsim 10^{10}$ and $z \lessapprox 2$. However, we also use galaxy shapes to break the spheroid/disc dichotomy, and detect evolution in the fraction of non Hubble sequence (asymmetric) galaxies. These galaxies increase in frequency with redshift at all stellar masses, and dominate the stellar mass over Hubble types at $z \gtrsim 2$ (Fig.~\ref{fig:cum}). Redshift evolution of galaxy rotational dynamics at a given stellar mass is also found, leading to evolution in the disc fraction if disc orbits are required to be sufficiently circular.

As in \citet{Clauwens17}, we infer that while mergers are important for the growth of spheroids, they do not drive the net destruction of discs. \citet{Clauwens17} found that mergers  contribute to both the destruction and growth of galaxy discs, resulting in a slight positive contribution to the mass in galaxy discs overall. In our taxonomy, mergers grow the mass in stellar spheroids at the expense of asymmetric systems, which are more prevalent at lower stellar mass. 

Despite the influence of mergers, we do find that stellar discs are steadily depleted by transformational processes, at a rate that is balanced by their replenishment through star formation at $z\sim0$ (Fig.~\ref{fig:specrate}). This depletion is thus attributed to secular processes. As a result, the stellar mass in discs is subdominant and gradually declining at lower redshifts, despite star formation proceeding almost exclusively ($>80$\%) within them. The secular destruction of discs supplements the growth of stellar spheroids. 

We have presented a study of how stellar structures form and evolve within the EAGLE simulations, given certain choices for how different structures are defined. However, a notable omission from this work is a direct comparison to observational data. Such a comparison is, of course, desirable to assess the validity of galaxy morphologies in EAGLE, and thus how applicable our findings are to the real Universe. In a future work we will exploit the virtual radiative transfer observations developed for EAGLE \citep[see][]{Camps16, Trayford17} to emulate how morphologies are measured observationally. This can be achieved for simulated populations using mock imaging \citep[e.g.][]{Snyder15, Dickinson18} or mock IFU data \citep[e.g.][]{Lagos17b}.  
 We hope that this study will provide a physical framework, useful to interpret observations and to provide more insight into the morphologies of real galaxies when paired with a direct observational comparison.

%% file: Appendices.tex
\appendix

\section{Mass and Size convergence}
\label{ap:conv}

\begin{figure*}
\includegraphics[width=0.95\textwidth]{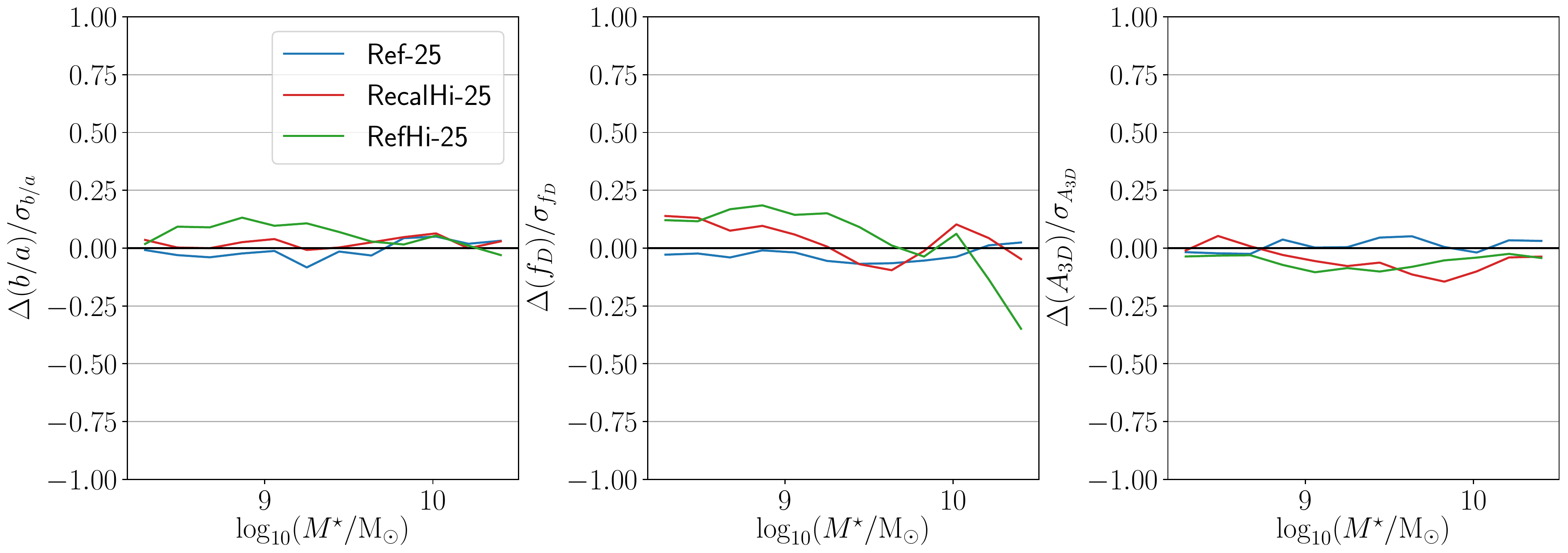}
\includegraphics[width=0.95\textwidth]{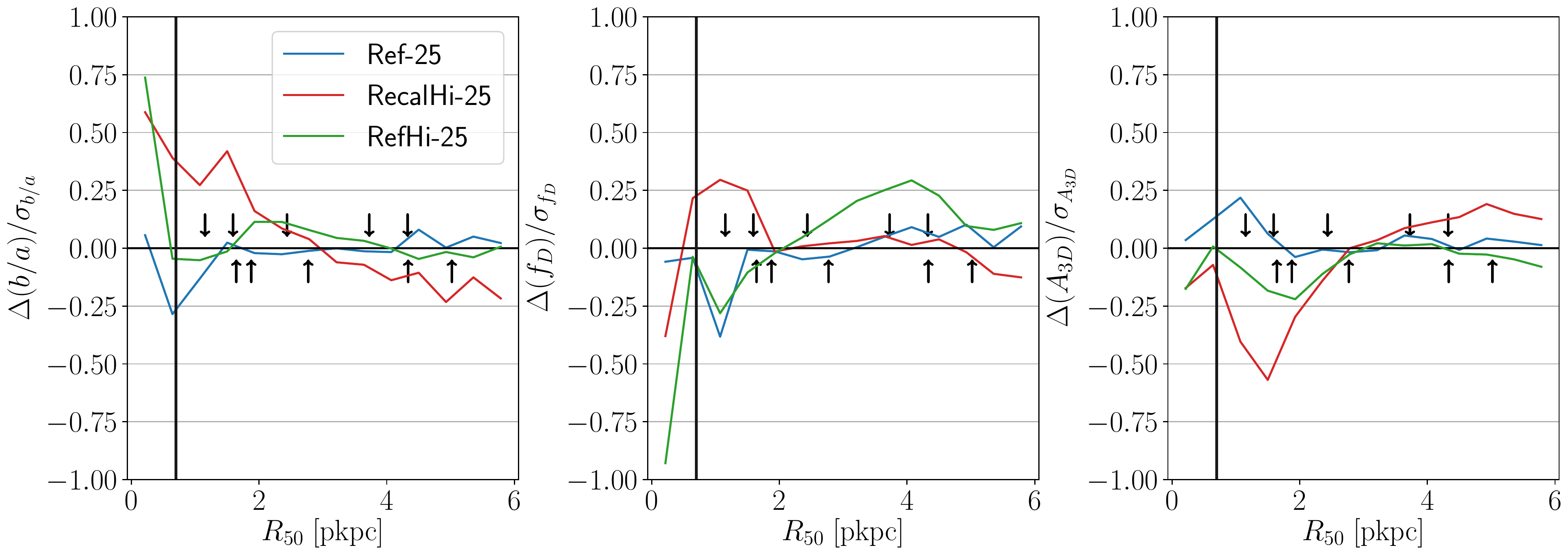}
    \caption{Stellar mass and size convergence with numerical resolution for three morphological metrics used in this work: $b/a$, $f_D$ and $A_{\rm 3D}$. Coloured lines represent the difference between the median values of each metric in a particular  25$^3$~Mpc$^3$ EAGLE volume from that of the Ref-100 simulation, as a function of galaxy stellar mass (top row) and size (bottom row). The Ref-100 and Ref-25 have the same fiducial resolution, while the resolution of RecalHi-25 and RefHi-25 is a factor 8 finer in mass and a factor 2 higher in length scale (see table~\ref{tab:sims}). This difference is normalised by the 16$^{\rm th}$-84$^{\rm th}$ percentile range of the metric in the Ref-100 volume. The 5, 10, 50, 90 and 95$^{\rm th}$ percentiles of the galaxy  size distribution are also indicated for all galaxies (up arrows) and high redshift galaxies ($z>1.5$, down arrows). The vertical lines in the bottom row represent the Plummer equivalent gravitational softening scale at fiducial resolution, 0.7~pkpc.}
\label{fig:conv}
\end{figure*}

To test numerical convergence, we consider both the effects of mass and spatial resolution, both of which are potentially important for measuring morphology. We test both \textit{strong} and \textit{weak} convergence properties\footnote{See S15 for a full discussion of these terms in the context of EAGLE.}. The former is achieved using the RefHi-25 run, which uses the fiducial model with resolution a factor 8 finer in mass and a factor 2 finer in length scale. The latter test uses RecHi-25, which matches the resolution of RefHi-25, but where the key parameters are recalibrated to reproduce the $z=0.1$ mass function at the new resolution. We test the convergence of three morphological metrics most relevant to this work: $b/a$, $f_D$ and $A_{\rm 3D}$.

Fig.~\ref{fig:conv} shows the difference in the median values of various metrics between the Ref-25, RefHi-25 and RecalHi-25 and the Ref-100 values as functions of stellar mass and stellar half-mass radius, scaled by the intrinsic 15.9-84.1 percentile range ($\sigma$) of the metric in the Ref100 volume. These plots include all $M^\star > 10^{8.5} {\rm M\odot}$ galaxies from all redshift outputs ($z \lesssim 6$) of each simulation. The mass dependence (top row) shows that the median values of each metric appear relatively well converged, remaining within 25\% of $\sigma$ over the mass range we consider. The weak convergence of $b/a$ and $A_{\rm 3D}$ are particularly good at low masses. The tendency of low-mass galaxies in the higher-resolution runs to have higher disc fractions \citep[as found by][]{Clauwens17} is recovered. 

Now we turn to the convergence properties of galaxy size between simulations. To describe the underlying distribution of galaxy sizes, we use arrows to indicate the (5, 10, 50, 90 and 95$^{\rm th}$) percentiles of the size distribution for all galaxies (up arrows) and high-redshift galaxies ($z>1.5$, down arrows). We also plot the Plummer equivalent gravitational smoothing scale at fiducial resolution ($\epsilon_{\rm s}$, 0.7~pkpc) as vertical lines. We see that there are considerable systematic variations for the most compact galaxies ($\lessapprox 1.5$~pkpc or 2$\epsilon_{\rm s}$). However, this only involves $\approx 5$\% of galaxies ($\approx 10$\% at high redshift). Regardless, it is important to be aware that resolution limitations affect the morphology of galaxies with sizes comparable to the gravitational smoothing.

\section{The asymmetry threshold}
\label{ap:pecs}

In order to study asymmetric galaxies, we must choose some criterion to define them. In this work we use a threshold value of the asymmetry parameter, $A_{\rm 3D}=0.2$, below which galaxies are classified as HS members and above which as asymmetric. This choice is somewhat \textit{ad-hoc}, and as such, we chose a cut between the high- and low-redshift peaks in the $A_{\rm 3D}$ distribution (Fig.~\ref{fig:a3ds}). 

\begin{figure}
\includegraphics[width=0.49\textwidth]{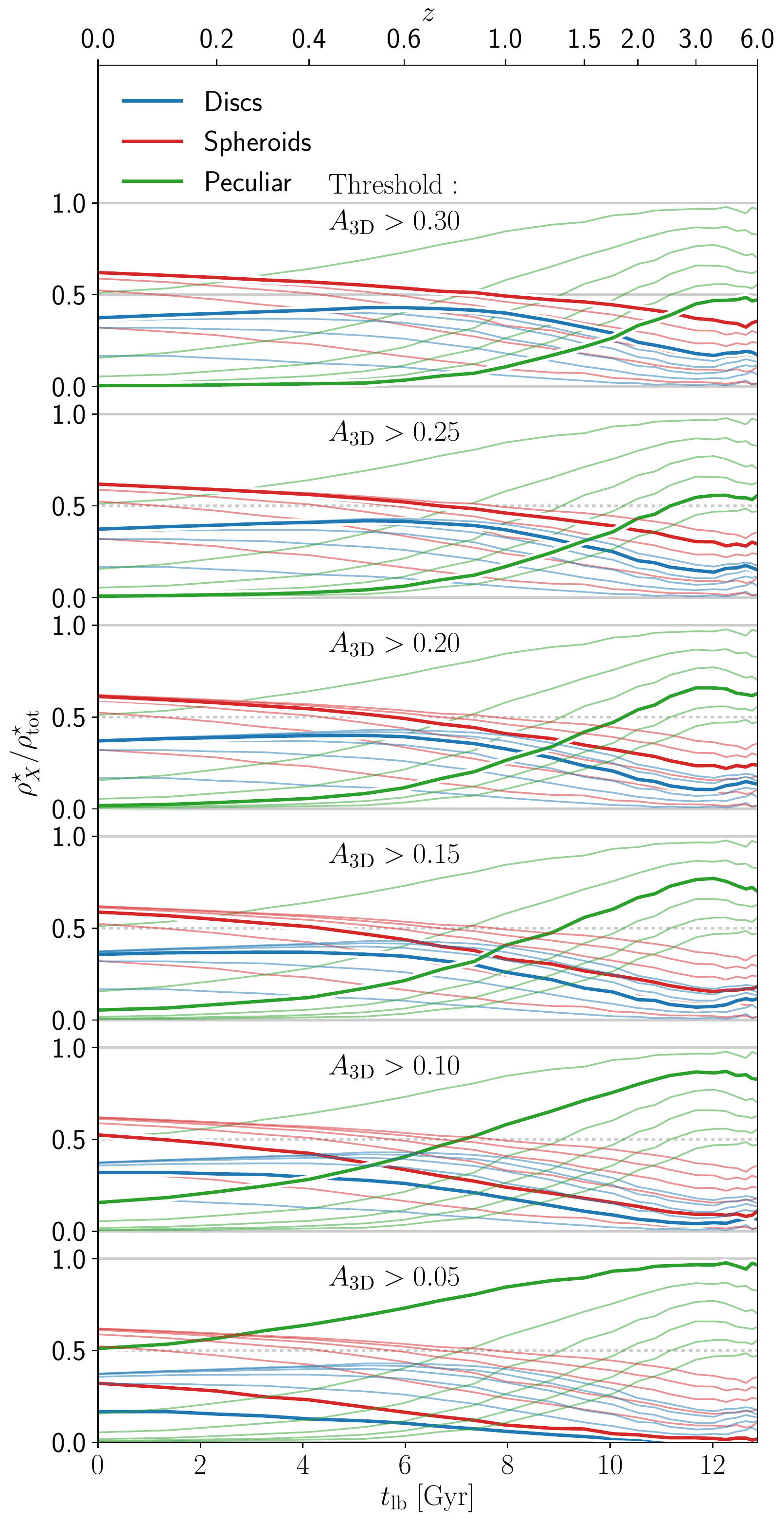}
    \caption{The fractional evolution of  disc, spheroid and asymmetric stellar structure contributions to the cosmic mass budget, with varying $A_{\rm 3D}$ thresholds to separate asymmetric and Hubble sequence galaxies. The threshold range is varied from 0.05 to 0.3 in steps of 0.05. Each sub-panel represents a given $A_{\rm 3D}$ cut, and shows the lines from all thresholds and morphological components for comparison, with the lines corresponding the appropriate threshold highlighted. We see that while absolute fractions vary, the generic evolutionary behaviour of the mass contributions is preserved.}
\label{fig:thresh}
\end{figure}

Changing this value will naturally change the absolute contributions of structures through cosmic time, and quantitative values such as the redshifts at which structures become dominant. Of course any classification of morphology is definition dependent, but our aim in this work is only to characterise the generic evolution of different structural types. Fig.~\ref{fig:thresh} demonstrates the robustness of this evolution with redshift.

In Fig.~\ref{fig:thresh} we show the evolving fractional stellar mass contributions of discs, spheroids and asymmetric structures for a number of threshold $A_{\rm 3D}$ values, from 0.05 to 0.3 in steps of 0.05. We see that while more stringent (higher) cut values naturally yield fewer galaxies that are asymmetric, the near exponential fall in the asymmetric fraction with cosmic time is preserved. Similarly, we see the gradually rising spheroid and disc fractions, with the disc contribution consistently peaking at $z\approx1$. From this we conclude that the generic evolution we describe in this work is relatively robust to the choice of asymmetry threshold.

\section{Influence of evolving demographics on gas circularities}
\label{ap:enviro}

\begin{figure*}
\includegraphics[width=0.99\textwidth]{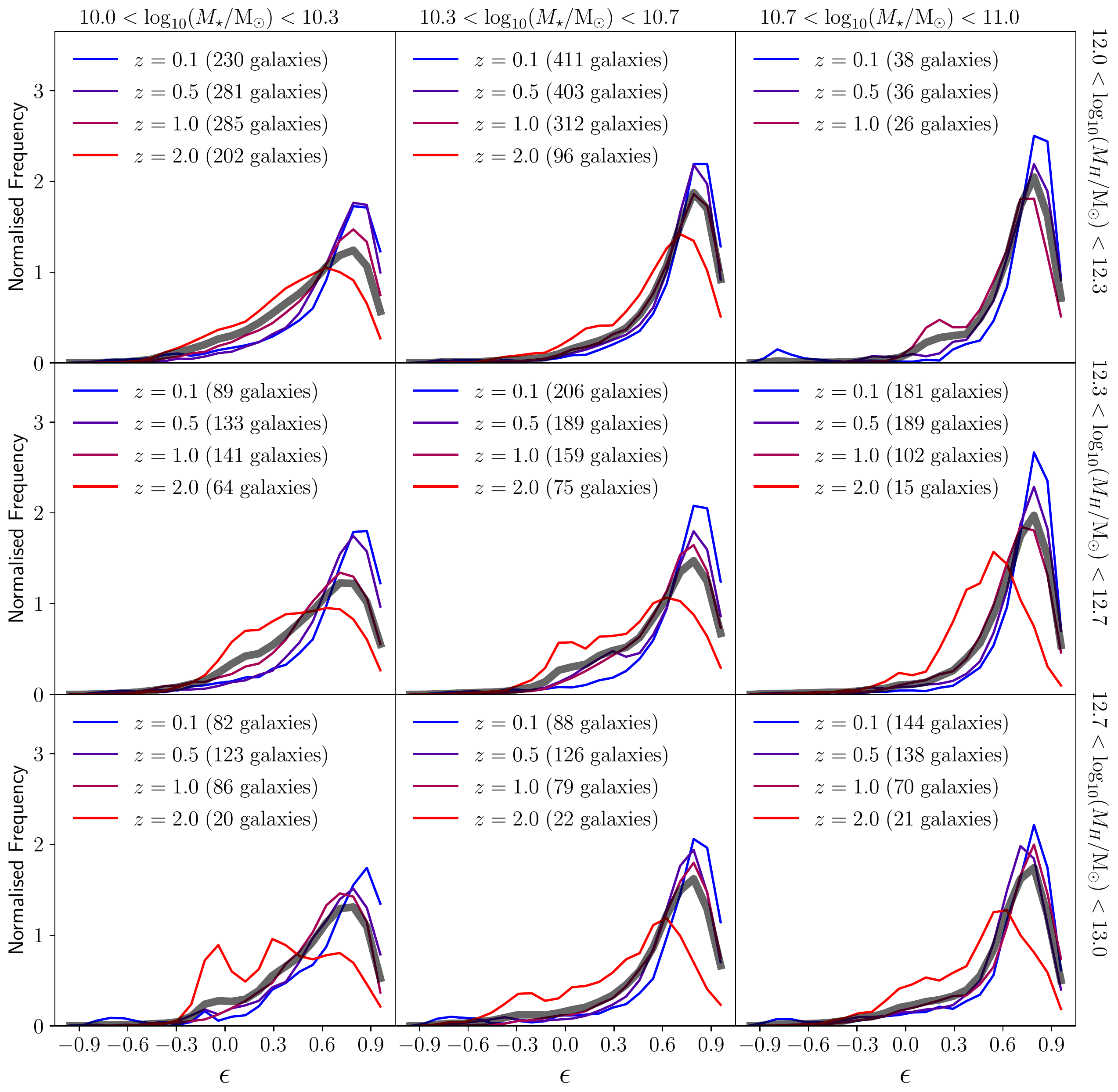}
    \caption{SFR-weighted circularity distributions for galaxies binned by stellar and halo mass. As in Fig.~\ref{fig:sfeps}, these distributions are measured at 4 distinct redshifts. The overall distribution (combining all 4 redshifts) is plotted as the grey translucent line to aid comparison between stellar and halo mass bins. We also indicate how many separate galaxies contribute to each distribution. Comparing distributions for a given redshift between stellar and halo mass bins reveals systematic differences, showing that galaxy demographics contribute to the overall evolution in gas circularity. However, comparable differences are found between distributions at fixed stellar and halo mass, suggesting that demographics alone cannot explain the entire circularity evolution.}
\label{fig:enviro}
\end{figure*}

We inspected the evolving kinematics of star-forming gas in section~\ref{sec:micro}, and found strong redshift trends in the SFR-weighted circularity distribution. However, it is unclear from this alone whether these trends can be ascribed to actual evolution in the nature of star-forming gas globally, or merely evolution in the galaxy population used to construct these distributions. 

We attempt to isolate these effects in Fig.~\ref{fig:enviro}, where we plot the SFR-weighted circularity distributions at each redshift constructed from galaxies binned by both stellar and halo mass. The distributions are plotted in each bin as in Fig.~\ref{fig:sfeps}, with a composite distribution plotted in grey. We note that `noise' in the distributions is not limited by particle counts, but rather by the galaxy counts that contribute to each distribution. This is because individual galaxies have distinct circularity distributions. We quote the number of galaxies contributing to each distribution in the legend.

We can look at the influence of galaxy mass and environment by comparing the same distribution between bins. Comparing the composite distributions between bins shows some variation. In particular, a systematic variation can be seen in the lowest redshift distributions, where for bins of higher stellar mass at a given halo mass, star-forming gas has higher circularity. However, this variation alone cannot explain the overall redshift trend, as $\epsilon$ distributions at different redshifts show comparable differences for fixed stellar and halo mass. This suggests that actual evolution in the properties of star-forming gas contributes to the trends between SF gas circularity and redshift, rather than merely the changing demographics of galaxies sampled at each redshift. 

We note that observations point to clear trends between stellar morphology and galaxy mass and environment. This is not necessarily inconsistent with what we find for the gas; it merely shows that the kinematics of star-forming gas are similar between galaxies of differing mass and environment where they occur at a given redshift. In addition, we find significant mass and environmental variation in morphology using the stellar circularity distributions, which are not plotted here. We leave discussion of this to future work.